\documentclass[debug]{rmaa}

\usepackage{paralist}

\usepackage{psfrag,color}

\usepackage[latin1]{inputenc}

\title{C,$T_1$,$T_2$: A complementary method to detect Multiple Populations with the Washington filter system}

\author{
  Heinz Frelijj,\altaffilmark{1} 
  Douglas Geisler,\altaffilmark{1,2,3}
  Sandro Villanova\altaffilmark{1}
  Cesar Munoz\altaffilmark{3,2,1}}

\altaffiltext{1}{Departamento de Astronomia, Casilla 160-C, Universidad de Concepcion, Concepcion, Chile.}
\altaffiltext{2}{Instituto de Investigacion Multidisciplinario en Ciencia y Tecnologia, Universidad de La Serena. Avenida Raul Bitran S/N, La Serena, Chile.}
\altaffiltext{3}{Departamento de Astronomia, Facultad de Ciencias, Universidad de La Serena. Av. Juan Cisternas 1200, La Serena, Chile.}

\shortauthor{Frelijj, Geisler, Villanova \& Munoz}
\shorttitle{C T1 T2: A complementary method to detect MPs}

\fulladdresses{
\item Heinz Frelijj: Departamento de Astronomia, Casilla 160-C, Universidad de Concepcion, Chile. (heinzfrelijj@gmail.com).
\item Douglas Geisler: Departamento de Astronomia, Casilla 160-C, Universidad de Concepcion, Chile; Instituto de Investigacion Multidisciplinario en Ciencia y Tecnologia, Universidad de La Serena. Avenida Raul Bitran S/N, La Serena, Chile; Departamento de Astronomia, Facultad de Ciencias, Universidad de La Serena. Av. Juan Cisternas 1200, La Serena, Chile.
\item Sandro Villanova: Departamento de Astronomia, Casilla 160-C, Universidad de Concepcion, Chile.
\item Cesar Munoz: Departamento de Astronomia, Facultad de Ciencias, Universidad de La Serena. Av. Juan Cisternas 1200, La Serena, Chile; Instituto de Investigacion Multidisciplinario en Ciencia y Tecnologia, Universidad de La Serena. Avenida Raul Bitran S/N, La Serena, Chile; Departamento de Astronomia, Casilla 160-C, Universidad de Concepcion, Chile. (cesar.alejandro.munoz.g@gmail.com)}

\listofauthors{Heinz Frelijj, Douglas Geisler, Sandro Villanova \& Cesar Munoz}
\indexauthor{Frelijj, H.}
\indexauthor{Geisler, D.}
\indexauthor{Villanova, S.}
\indexauthor{Munoz,C.}

\abstract{In this research we test the ability of a three Washington filter combination, $(C-T_1)-(T_1-T_2)$, compared with that of the traditional $C-T_1$ color to find multiple populations on two globular clusters: NGC 7099 and NGC 1851, types I and II Globular clusters, respectively. Our improved photometry and membership selection, now using Gaia proper motions, finds that the second population stars are more centrally concentrated than first population stars, as expected and contrary to our previous findings for NGC 7099. We find that multiple populations are more easily detected in both clusters using the new $(C-T_1)-(T_1-T_2)$ color, although $C-T_1$ conserves the best width/error ratio. We also search for differences of both colors while splitting the red-RGB and the blue-RGB in NGC 1851, but no significant improvement was found.}

\resumen{}

\addkeyword{errata}
\addkeyword{globular clusters: individual: NGC 7099}
\addkeyword{globular clusters: individual: NGC 1851}
\addkeyword{Hertzsprung$-$Russell and colour$-$magnitude diagrams}
\addkeyword{stars: imaging}
\addkeyword{techniques: photometric}

\begin{document}
\maketitle

\section{Introduction}
\label{sec:intro}

Multiple Populations (MPs) are now recognized as an essential characteristic of almost all Globular Clusters (GCs). Historically, \citet{Cannon1973} almost 50 years ago found an exceptionally large scatter in the Red Giant Branch(RGB) of Omega Cen. Thirty years later \citet{Bedin2004} found two primary sequences not only in the color of the RGB of Omega Cen, but also in the Sub Giant Branch (SGB) and Main Sequence(MS), giving the first detailed photometric glimpse of what would become known as MPs.
 But it was not until \citet{Carretta2009} realized a heroic high resolution spectroscopic study of thousands of stars in a total of 19 GCs, finding chemical inhomogeneities in all of them, that MPs began to be considered as an intrinsic characteristic of GCs. Subsequently, \citet{Carretta2010} in fact suggested a new defintion of a GC: Stellar systems showing anti-correlations among the abundances of light-elements, whose main and most widespread example is the Na-O anticorrelation. This major study would be complemented 6 years later with that of \citet{Piotto2015}, who conducted the HST GC UV Legacy survey using an improved photometric method employing the UV/blue WFC3/UVIS filters F275W, F336W, and F438W , best known as the "magic trio", to characterize MPs in 57 GCs, showing that they all possess MPs and proving photometry with appropriate filters as an excellent method to detect MPs. The advantages of photometry over spectroscopy of course are the ability to investigate MPs in a much larger sample in a given GC with a much smaller telescope than typically possible with high resolution spectra. Two years later, \citet{Milone2017} divided these 57 GCs into type I(GCs whose stars separate in two distinct groups, identified as first(1P) and second(2P) populations) and type II clusters (those GCs where the 1P and/or the 2P sequences appear to be split and include an additional group of redder stars in the chromosome map. Type II GCs also exhibit multiple SGBs in purely optical CMDs).
 
 The most important conclusion is that virtually all the GC show MPs. But some clusters seemed to be the exception: IC4499 \citep{Walker2011}, E3 \citep{Salinas2015}, Terzan 7 \citep{Tautvaivien2004} and especially Ruprecht 106 \citep{Villanova2013,Frelijj2021} are the best examples (altough certain HST studies put in doubt some of them \citep{Dotter2018,Dalessandro2018}). Thus, every cluster has to be carefully studied to determine whether it has MPs or not, and study its characteristics, as the UV Legacy survey has demonstrated that every GC is unique in its MP behavior.
 
Various scenarios for the origin of MPs have been proposed: Asymptotic Giant Branch scenarios like \citet{Dercole2008}, Fast Rotating Massive Stars scenarios like \citet{Decressin2007} and even a scenario that did not invoke multiple epochs of star-formation \citep{Bastian2013}. But currently none of them satisfies all the observational evidence \citep{Renzini2015,Bastian2018}(although newer models, like \citet{elmergreen2017,gieles2018} and \citet{parmentier2022} are getting closer).

The previous studies mentioned above have proven photometry to be a very good way to search for MPs, because, while it cannot provide the detailed abundances of spectroscopy, it allows the measurement of a much larger sample of stars simultaneously and to much fainter absolute magnitudes. \citet{Sbor2011} produced synthetic spectra of two otherwise identical GC giants, one being a 1P star with normal chemical abundances of the light elements and the other is a 2P star with enhanced He, N and Na and depleted C and O, as observed in many spectroscopic studies. The study shows that significant differences in flux between the two spectra exist and are related to the various CN, CH, NH and OH bands, particularly in the UB/blue part of the spectrum. 

Some photometric bands, concentrated in the blue-uv portion of the spectrum, are specially sensitive to these bands. The best known are the already mentioned "magic trio" of filters used in \citet{Piotto2015}, consisting of three HST UVIS/WFC3 filters: F275W, F336W and F438W. The combination of these sensitive filters led to colors maximizing the separation of the different populations of stars. Actually, most blue/UV filters are capable of uncovering MPs. However, although most such filters,
such as $U_{Johnson-Cousins}$ and $u'_{SDSS}$, detect MPs quite well, they require long exposure times due to their relatively narrow band and/or low efficiency.
 
The Washington filter system was designed by \citet{Canterna1976} originally to derive a photometric temperature (from the $T_1$ and $T_2$ filters, very similar to $(RI)_{KC}$), as well as a metallicity index (from the M filter) for G and K giants. However, at the time, CN and CH variations were being discovered in GCs and it was felt prudent to include another filter that would be sensitive to such variations independent from metallicity effects, and thus the C ("Carbon") filter was added. The Washington C filter is a blue-UV filter, with $\lambda_{eff}=$3982\AA \ and $\Delta\lambda=$1070\AA \ \citep{Bessel2005}. This broadband allows it to encompass 3 CN-Bands and one NH-Band, as well as the CH band. Because of its efficiency, it should be sensitive to MPs in considerably less exposure time than other, more narrow-band, filters. These 2 characteristics make the Washington C filter a good option for detection of MPs. Indeed, the HST WFC3/UVIS instrument includes a C-like filter: F390W.
 
 Initial efforts investigating the possibility of uncovering MPs from the ground with the Washington system used the $C-T_1$ color, obtaining good results \citep{Cummings2014,Frelijj2017}. These results, although not as accurate as HST data, present an attractive alternative, based on small ground-based telescopes. Our aim here is to investigate if there might be an even better Washington color for detecting MPs, involving the addition of the T2 filter, which maintains some MP sensitivity (see Figure \ref{synth}).
 
 This paper is organized as follows:\\
 
 In section II we present the data used, how it was obtained and reduced.\\
 Section III describes the results using the the new method and compares these with results from the initial technique. We also analyze the results.\\
 Section IV contains a summary of the paper.

\section{Data}
\label{sec:data}
\subsection{Observations}
The data consist of 46 images, 23 of NGC 7099 and 23 of NGC 1851. They were obtained from 2 telescopes, the 1-meter Swope telescope from Las Campanas Observatory, Chile; and the 4m SOAR telescope on Cerro Pachon, Chile.
 The filters selected for this work were the Washington C filter \citep{Canterna1976}, and the filters $R_{KC}$ and $I_{KC}$ in replacement of the Washington filters $T_1$ and $T_2$ since \citet{Geisler1996} demonstrated that the $R_{KC}$ filter is a more efficient substitute for $T_1$ and the $T_2$ filter is almost identical to $I_{KC}$ \citep{Canterna1976,Geisler1996}.
 For NGC 7099 we used the same images from \citet[hereafter F17]{Frelijj2017}, only dropping 1 medium and 2 long C exposures from the Swope Telescope in order to decrease the average seeing. The airmasses vary between 1.0-1.4 while the FWHM is 0.9''-1.7'' for the Swope images and 0.39''-0.54'' for SOAR images.
 For NGC 1851 we took the images used in \citet[hereafter C14]{Cummings2014} but discarded 3 long, poor-seeing C exposures from the Swope telescope, added 1 short exposure from Swope for C,R \& I, and added 2 short and 2 long exposures from SOAR in C. The airmasses vary between 1.0-1.5 while the FWHM is 0.9''-1.58'' for the Swope images and 0.49''-0.52'' for the SOAR images. All nights appeared photometric visually.

Table~\ref{images } gives the details of the exposures.

\begin{table}
	\centering
    \caption{NGC 1851 \& NGC 7099: Images} \label{images }
	\begin{tabular}{ccccccc} 
		\toprule
		\midrule
		& & & NGC 7099 & &\\
		\midrule
		 & & Swope & & SOAR\\
		\midrule
		C & 1(30s) & 2(300s) & 4(1200s) & 4(10s) & 2(300s)\\
		R & 1(10s) & 1(100s) & 3(400s) & - & -\\
		I & 1(10s) & 1(300s) & 3(1200s) & - & -\\
		\toprule
		\midrule
		& & & NGC 1851 & &\\
		\midrule
		 & & Swope & & SOAR\\
		\midrule
		C & 1(30s) & 1(300s) & 7(1200s) & 2(10s) & 2(300s)\\
		R & 1(10s) & 1(100s) & 3(400s) & - & - &\\
		I & 1(10s) & 1(300s) & 3(1200s) & - & - &\\
		\midrule
		\bottomrule
	
	\end{tabular}
\end{table}

The Swope images were observed with a CCD (SiTe3) of 2048x3150 pixels at 0.435 ''/pix and a field of view of 14.9 x 22.8 arc minutes. The SOAR detector (SOI) consists of a total of 4096 x 4096 pixels at 0.1534''/pix (0.0767 ''/pix binned 2x2) and a field of view of 5.26 x 5.26 arc minutes, divided into two CCDs with two amplifiers each resulting in 4 columns of 1024x4096 pixels.
 
 \subsection{Processing and Reduction}
IRAF \footnote{IRAF is distributed by the National Optical Astronomy Observatory, which is operated by the Association of Universities for Research in Astronomy (AURA) under a cooperative agreement with the National Science Foundation.} and its standard tasks were used to process all the photometric data.
  A linearity correction \citep{Hamuy2006} was applied to all the Swope (SiTe3) images in order to increase the range of unsaturated stars. 
  DAOPHOT \citep{Stetson1987} and its suite of tasks were used to perform the photometry in both clusters since it was specially developed to work on crowded fields. A first PSF was determined in each single image by taking the $\sim$200 brightest unsaturated and more isolated stars. These stars were refined subtracting all their detected neighbours to determine a second and more precise PSF that was refined a third time by eye, thus removing all PSF-stars with bad subtracted neighbours. This refined PSF determined in each image was applied to carry out a PSF photometry three successive times through the tasks FIND, PHOT and ALLSTAR. Due to the large pixel scale of the SiTe3 detector (0.435 ''/pix), we decided to repeat the technique used in C14, setting in DAOPHOT and ALLFRAME a fitting radius 0.4px smaller than the FWHM measured, for all the Swope images with a FWHM smaller than 3 px in order to avoid photometric errors due to "square stars".
   We experimented with different ALLFRAME \citep{Stetson1994} methods based on the procedures from C14 and F17, and found that the best photometry is obtained in the following way: \\
   -First, applying the cuts used in C14, that consists in removing all the stars with errors higher than 0.15, chi-squared greater than 2.5, absolute sharpness value greater than 1 (1.5 for C filter) and magnitudes above the point (determined for each image by looking in the plot magnitude vs error) where the stars begin to be affected by the nonlinearity of the detector.\\
  -Second, using DAOMATCH and DAOMASTER to match all the images to create a single starlist that will be given to ALLFRAME to perform PSF-Photometry in all the images simultaneously.\\
  -Third, using DAOMATCH and DAOMASTER to match the catalogs given by ALLFRAME, first combining the images with the same time exposure and filter, and then all the resulting catalogs of each filter to get a robust intensity-weighted mean instrumental magnitude, using the medium exposure as a reference image since it maximizes the number of stars in common with both short and long exposures, facilitating the match. \\
  -Finally, use DAOMATCH and DAOMASTER again to generate a full catalog with all the stars found in at least 2 of the 3 filters. The R filter was used as reference filter since its wavelength response lies between the C and I filters and because it produces the deepest images. 

 Aperture corrections were determined taking the brightest and unsaturated stars from the entire field comparing their PSF photometry to their aperture photometry. No spatial dependence was found in any filter for both clusters.
 
 The instrumental magnitudes of NGC 7099 were transformed to the standard Washington system using the standard coefficients obtained in F17. The RMS for each filter is 0.038(C), 0.022(R) and 0.027(I). However, for NGC 1851, the standard coefficients from C14 caused an offset of $\sim$ 0.15 to the red in the RGB with respect to the CMD from C14, probably due to the addition of the new images, so we decided to calibrate calculating the difference between our instrumental magnitudes and the standard magnitudes of the CMD from C14 for each star using the following formulae: 
  $$ C = (c-r)*m_1 + n_1 + c$$ 
  $$T1 = (c-r)*m_2 + n_2 + r$$
  $$T2 = (r-i)*m_3 + n_3 + i$$
  
  where C, T1 and T2 are our calibrated magnitudes, m is the slope, n is the y-intercept of the line and c, r and i our instrumental magnitudes. The resulting calibrated magnitudes are very similar to those from C14.

  According to \citet{Bonatto2013} NGC 1851 has a 
  mean differential reddening of $\langle\delta E(B-V)\rangle=$ 0.025 $\pm$ 0.01 while NGC 7099 has a 
  mean differential reddening of $\langle\delta E(B-V)\rangle=$ 0.03 $\pm$ 0.01. Taking into account the relation from \citet{Geisler1991} E(C-T1) = 1.966(B-V) we obtain a reddening of E(C-T1)=0.049 for NGC 1851 and 0.059 for NGC 7099\footnote{\citet{Bonatto2013} say that differential reddening values lower than 0.04 may be related to zero-point variations.}. We consider these numbers small enough to be negligible, so reddening corrections are not needed. In particular, in this work we are only interested in differential effects between possible different MPs and not absolute effects.
  
 Finally, a World Coordinate System (WCS) was calculated in both NGC 1851 and NGC 7099 catalogues using 12 stars well distributed along the field to transform the x/y coordinates to RA/Dec(J2000) using the xy2sky task from WCSTools.
 
 \begin{table*}
\centering
\setlength{\tabnotewidth}{1.455\columnwidth}
\tablecols{11}
\setlength{\tabcolsep}{\tabcolsep}
\begin{changemargin}{-2.8cm}{-2.8cm}
\caption{NGC 1851 \& NGC 7099: Catalog example} \label{grandmaster }
\begin{tabular}{lllllllllll}
\toprule
ID & RA(J2000) & DEC(J2000) & X & Y & Rad & C & eC & dC & mC & nC\\
\midrule
324 & 78.684240068 & -40.042003974 & 161.219 & 1827.654 & 999.03 & 20.3654 & 0.0082 & 0.0033 & 0.0082 & 3\\
334 & 78.683674458 & -40.043001613 & 164.868 & 1835.867 & 995.23 & 20.7216 & 0.0101 & 0.0016 & 0.0101 & 3\\   
347 & 78.683039621 & -40.047603550 & 169.203 & 1873.867 & 991.08 & 20.8262 & 0.0113 & 0.0010 & 0.0113 & 3\\ 
\bottomrule
\tabnotetext{a}{The columns are: ID, RA and DEC coordinates (in degrees), X and Y coordinates (in px), radial distance of the star to the centre (in px), magnitude, psf-fitting error (internal error), dispersion (external error), higher value between internal and external error,  and the number of frames where the star was detected. (All of this for C, $T_1$ and $T_2$ but due to the lack of space this table shows only C)}
\end{tabular}
\end{changemargin}
\end{table*}  

 \subsection{Final sample selection}
 
  As mentioned in previous works, DAOMASTER gives two types of errors: the combined photometric measurement error output by ALLFRAME(internal error) and the $\sigma$ based directly on the observational scatter across multiple images(external error). We already proved in F17 through an ADDSTAR experiment that the external errors are better estimates of the real photometric error than internal errors, but for each star we take the largest of these two errors to avoid the fact that some stars detected in one single frame have error "0". These final errors appear in table \ref{grandmaster } as mC (We use C as an example for $T_1$ and $T_2$ too). We removed all the stars with errors greater than 0.1 in each filter, and colors were created from the remaining stars. The errors in colors are the square root of the quadratic sum of the final errors from each input magnitude. Radial cuts were applied to both clusters following the previous studies from C14 and F17. For NGC 7099 we removed all the stars from the center up to 80px(34.8'') radius while for NGC 1851 we cut up to 50px(21.75'') due to crowding and we left for both clusters a ring with an outer radius of 1000px(7.25').
 
\begin{figure*}
  	\includegraphics[width=0.5\columnwidth]{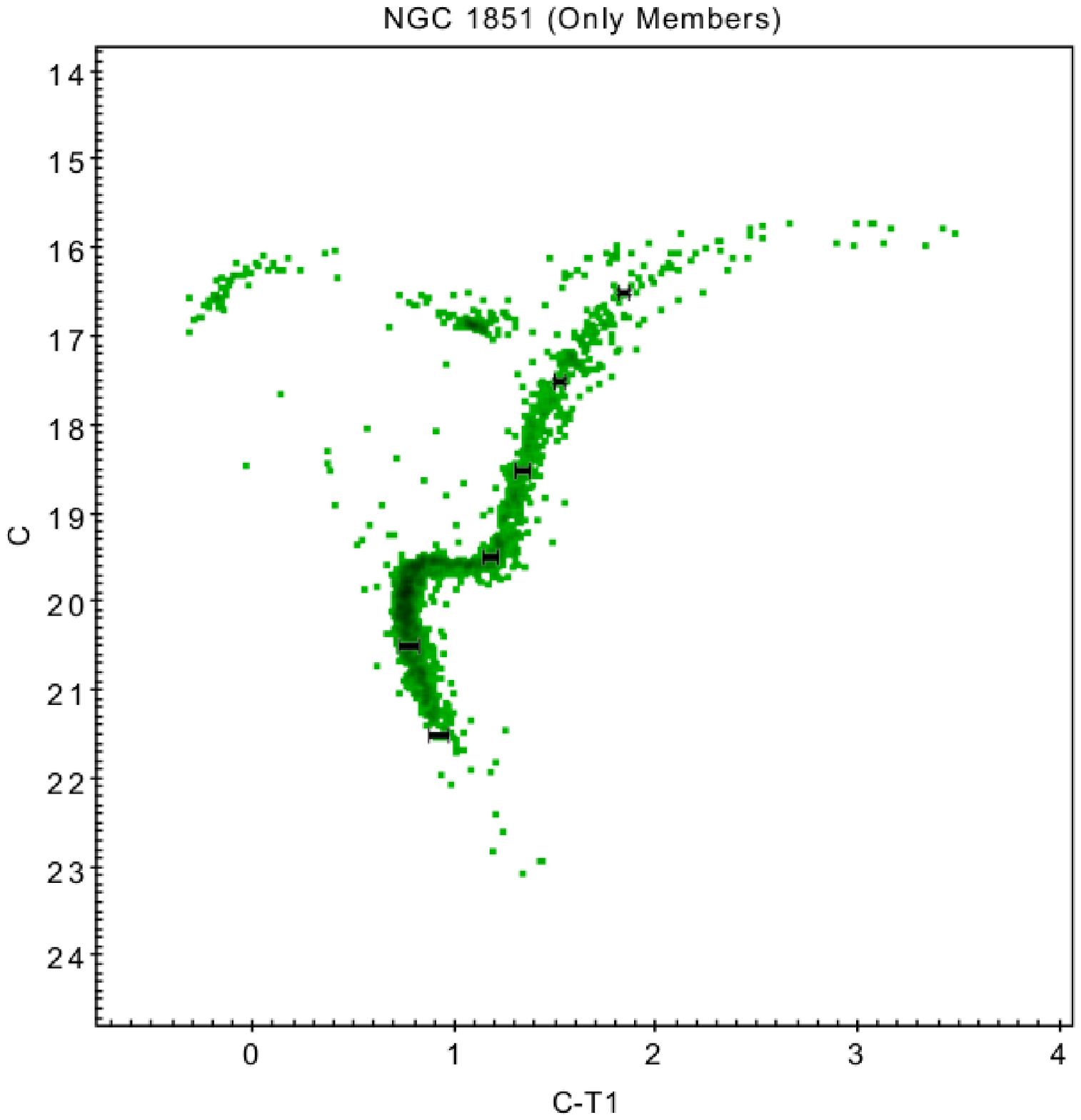}
  	\includegraphics[width=0.5\columnwidth]{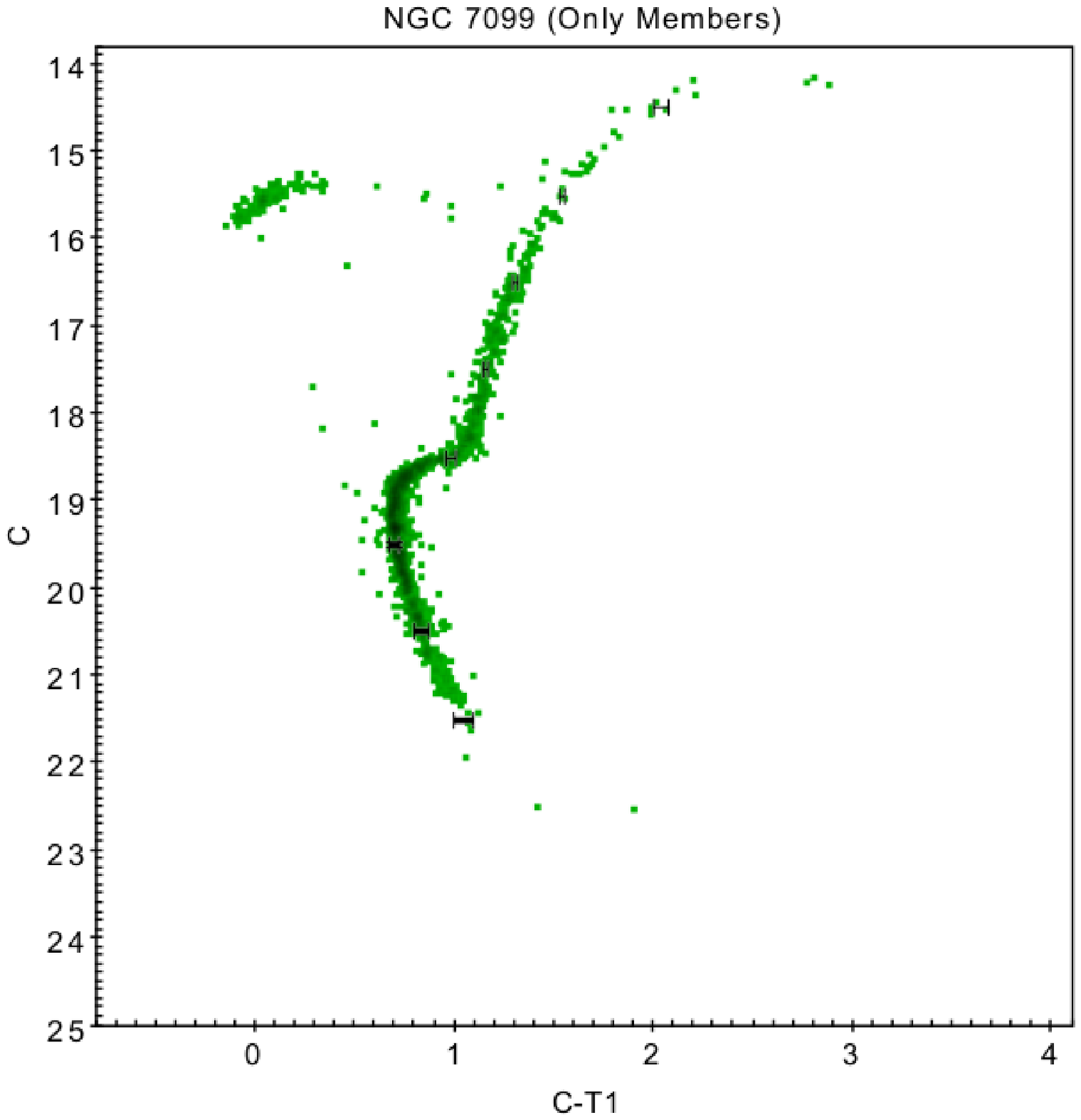}
	\includegraphics[width=0.5\columnwidth]{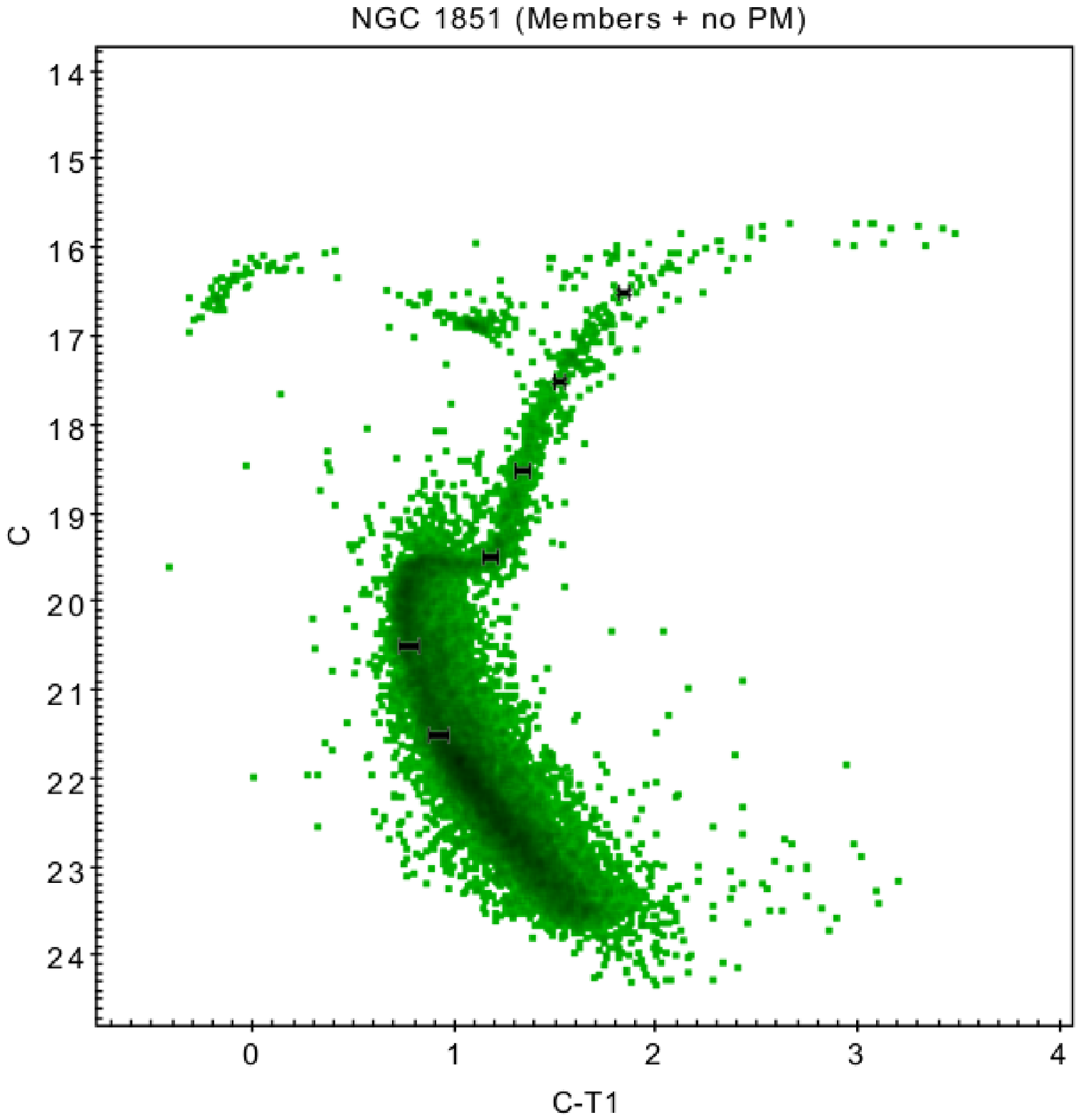}
	\includegraphics[width=0.5\columnwidth]{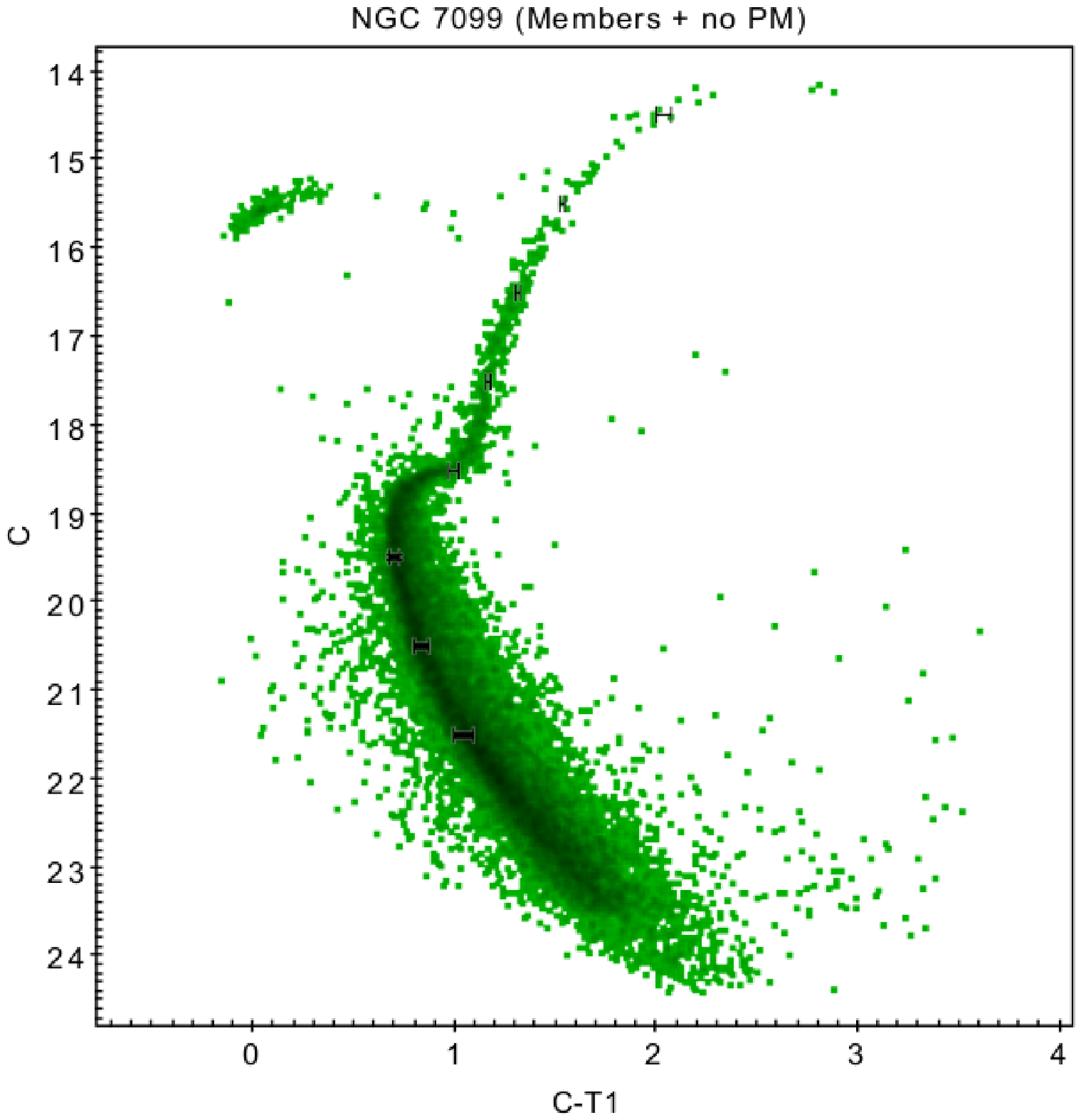}
    \caption[NGC 1851 \& NGC 7099: Definitive CMDs in C-T1 vs C]{Definitive CMDs in C-T1 vs C. Top left: NGC 1851 using only member stars according to PMs provided by GAIA. Bottom left: NGC 1851 using member stars according to PMs provided by GAIA plus stars without PMs detected. Top right: NGC 7099 using only member stars according to PMs provided by GAIA. Bottom right: NGC 7099 using member stars according to PMs provided by GAIA plus stars without PMs detected.}
    \label{gaia }
\end{figure*}

  Proper Motions (PM) provided by the \textit{Gaia} DR2 mission \citep{Gaia2016,Gaia2018} allowed to select (in a PM-RA vs PM-DEC plot) all the stars with PM similar to our cluster reproducing by hand the selection shown in the Baumgardt Globular Cluster database (3rd version) \footnote{https://people.smp.uq.edu.au/HolgerBaumgardt/globular/}, removing non-member stars and cleaning the CMD.
  From now we work with two kinds of catalog in each cluster, one catalog containing only member stars to ease the detection of different sequences or broadening in the clusters(Figures \ref{gaia }$_{Top}$), and a second catalog containing the same member stars plus all the stars that do not have a PM, aiming to have a deeper Main Sequence (Figures \ref{gaia }$_{Bottom}$).
  
Both catalogs show improvements with respect to their original papers: NGC 1851 is $\sim$1.5 mag deeper in C, and we can see better the double Sub-Giant Branch (SGB) mentioned in C14 and \citet{Han2009}, and that \citet{Milone2017} classified as a characteristic of type II GCs. For NGC 7099, since we discarded some bad seeing images, we have a CMD $\sim$1 mag deeper in C and $T_2$ and a narrower SGB.   

\section{The efficacy of the new color in detecting MPs}
 
\begin{figure*}
	\includegraphics[width=\textwidth]{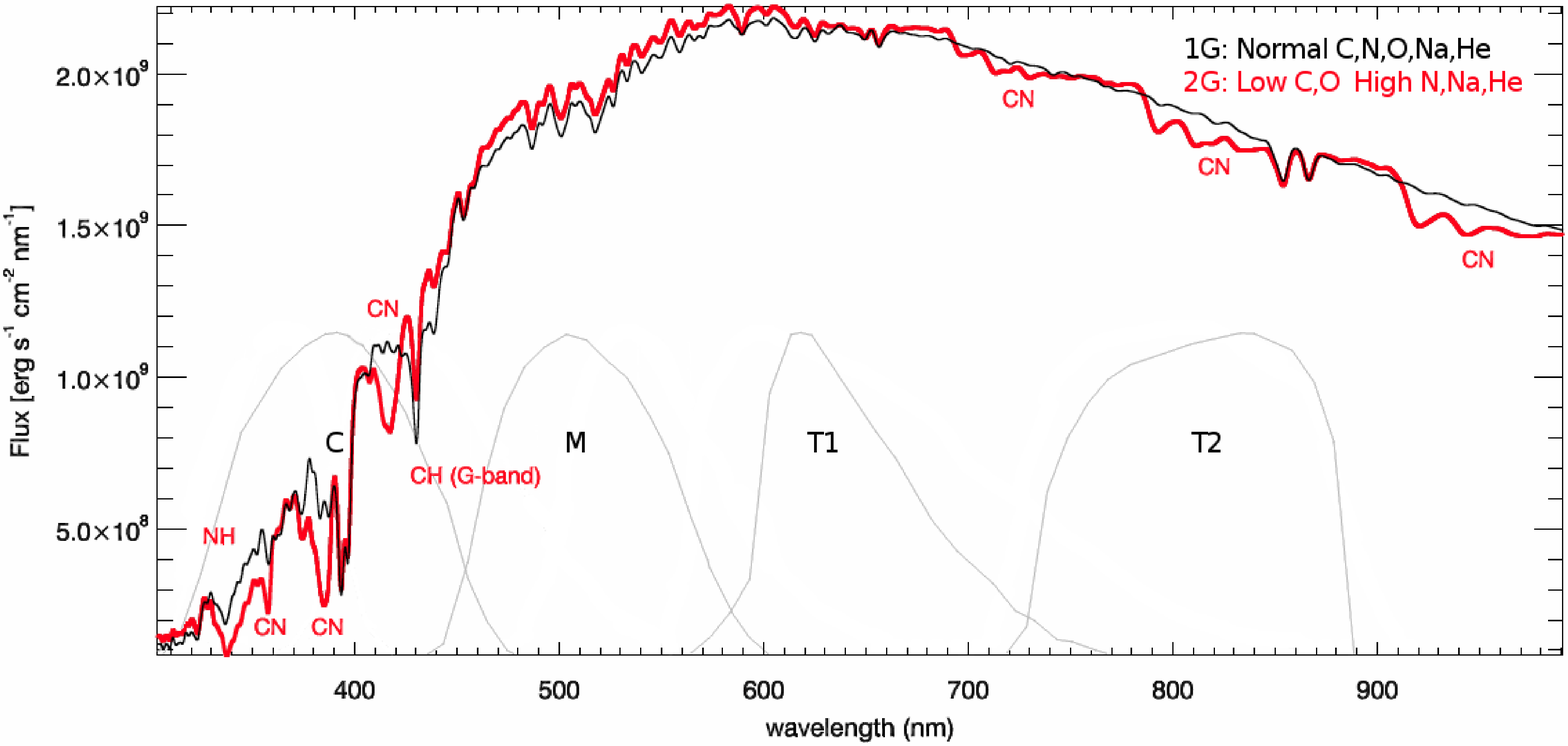}
    \caption{Comparison of the synthetic spectra from 1P(black) and 2P(red) stars. Illustrative Washington filter response curves are included.}
    \label{synth}
\end{figure*}

 As mentioned before, C14 and F17 proved the efficacy and efficiency of the Washington C filter to uncover MPs. This filter goes from the atmospheric cutoff at around 3300 \AA \ to beyond the G-band, thus covering 3 CN-bands, a NH-band and a CH-band. This can be seen in Figure \ref{synth}, that shows the comparison between the synthetic spectra of otherwise identical 1P and 2P stars made by \citet{Sbor2011} with the Washington filter response curves included. 
 
\begin{figure*}
  \centering
   	\includegraphics[width=0.392\columnwidth]{1851pm.eps}
  	\includegraphics[width=0.392\columnwidth]{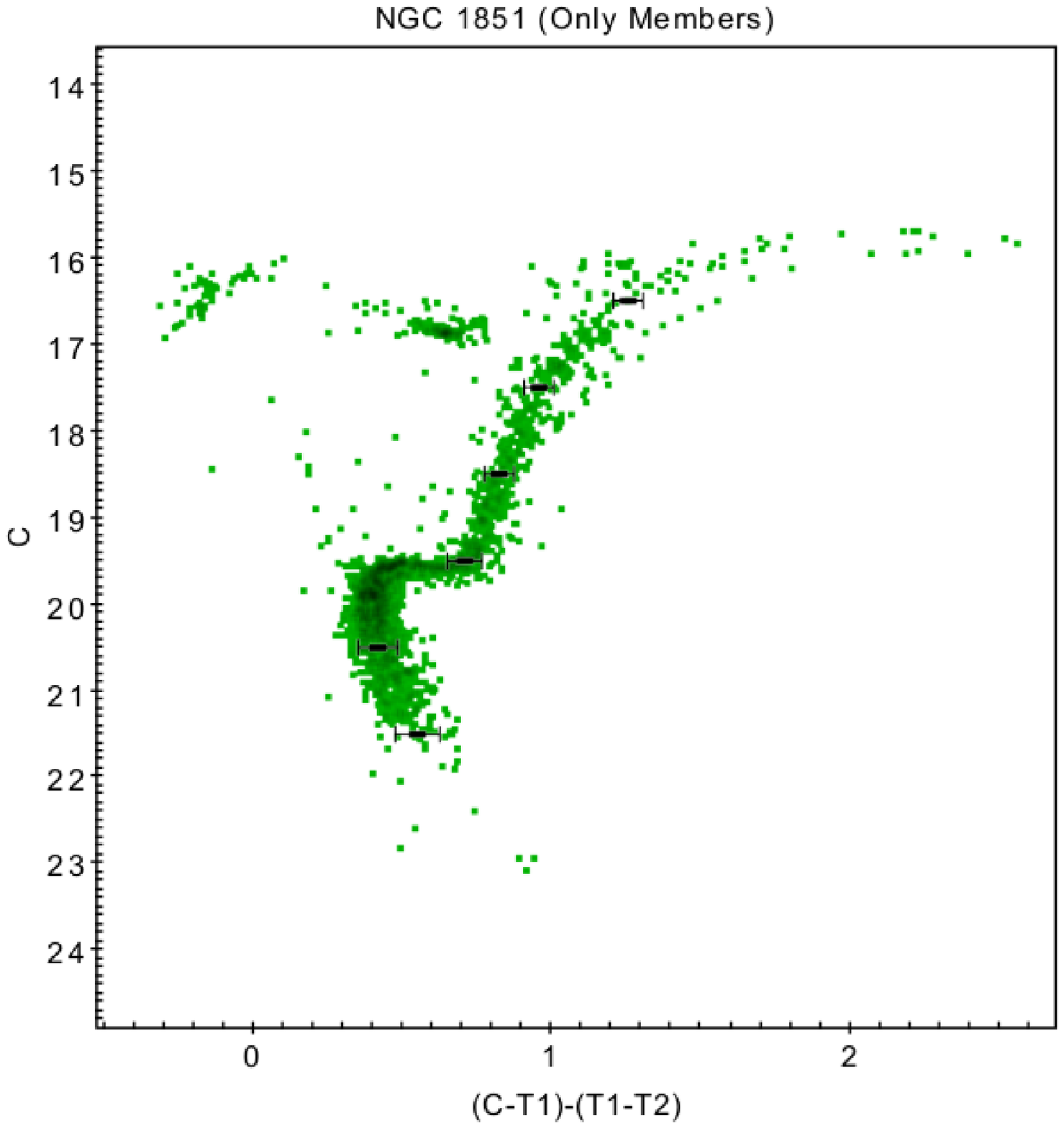}
  	
  	\includegraphics[width=0.392\columnwidth]{7099pm.eps}
  	\includegraphics[width=0.392\columnwidth]{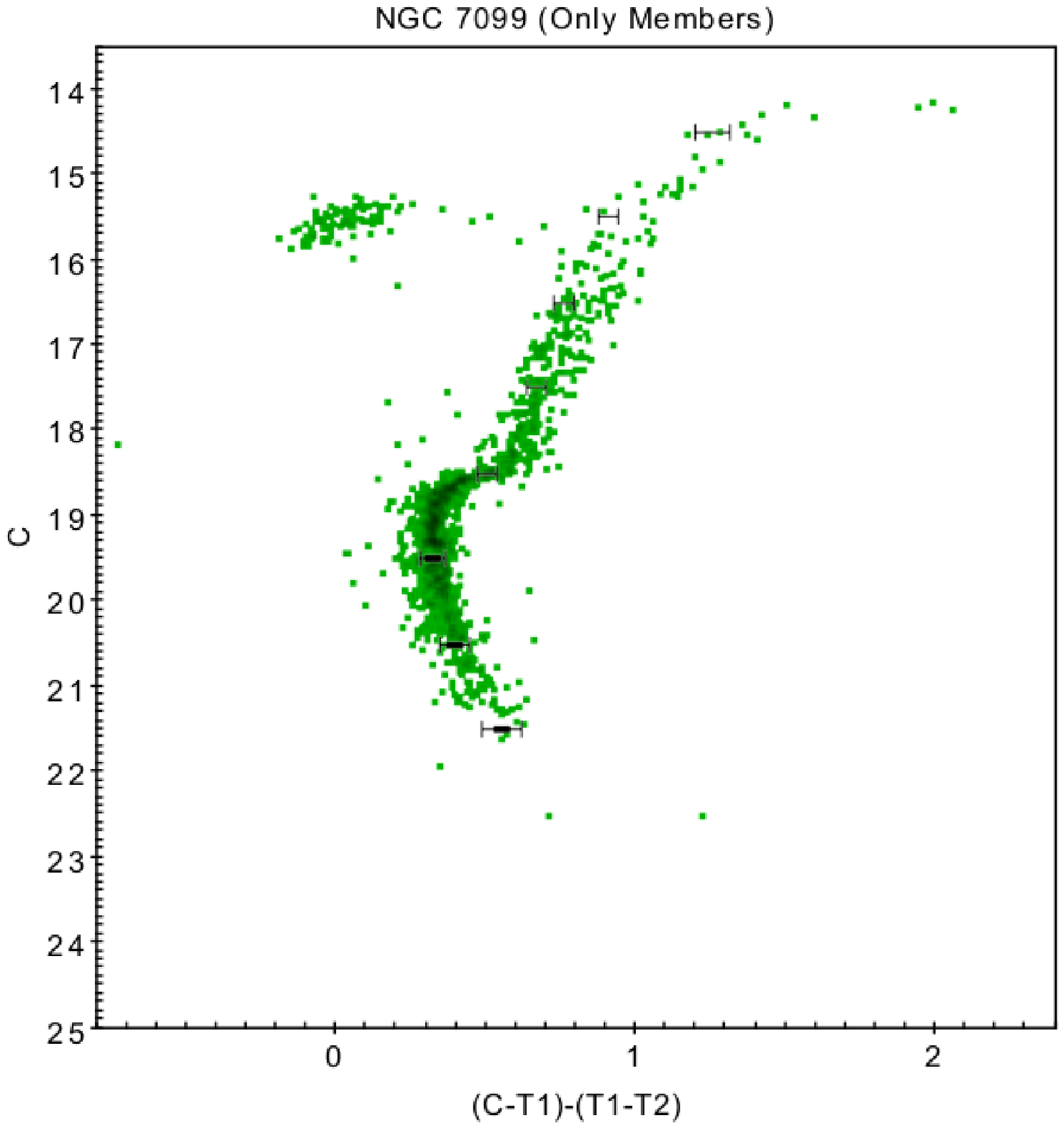}
  	
	\includegraphics[width=0.392\columnwidth]{1851pm+nopm.eps}
	\includegraphics[width=0.392\columnwidth]{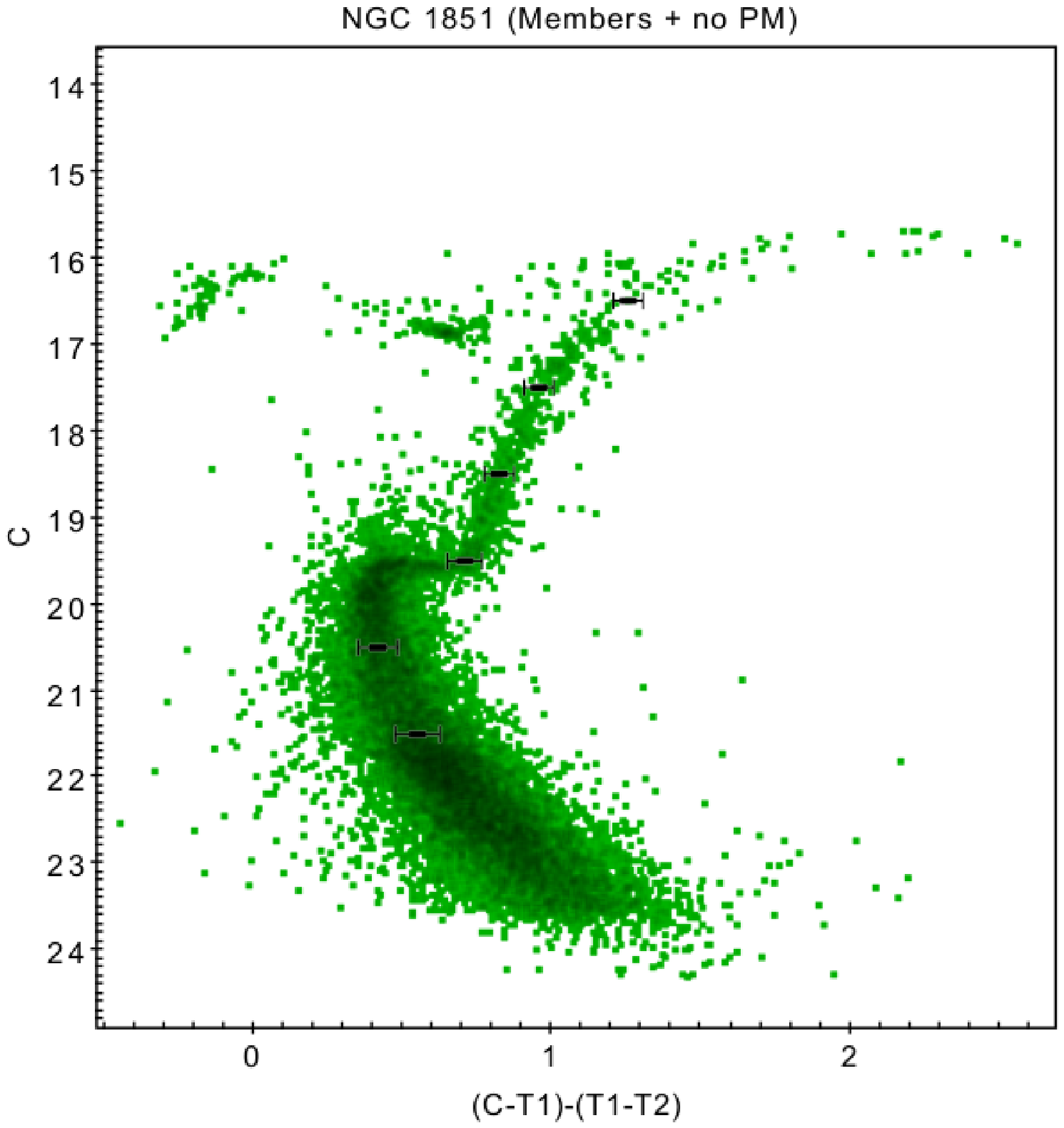}
	
	\includegraphics[width=0.392\columnwidth]{7099pm+nopm.eps}
	\includegraphics[width=0.392\columnwidth]{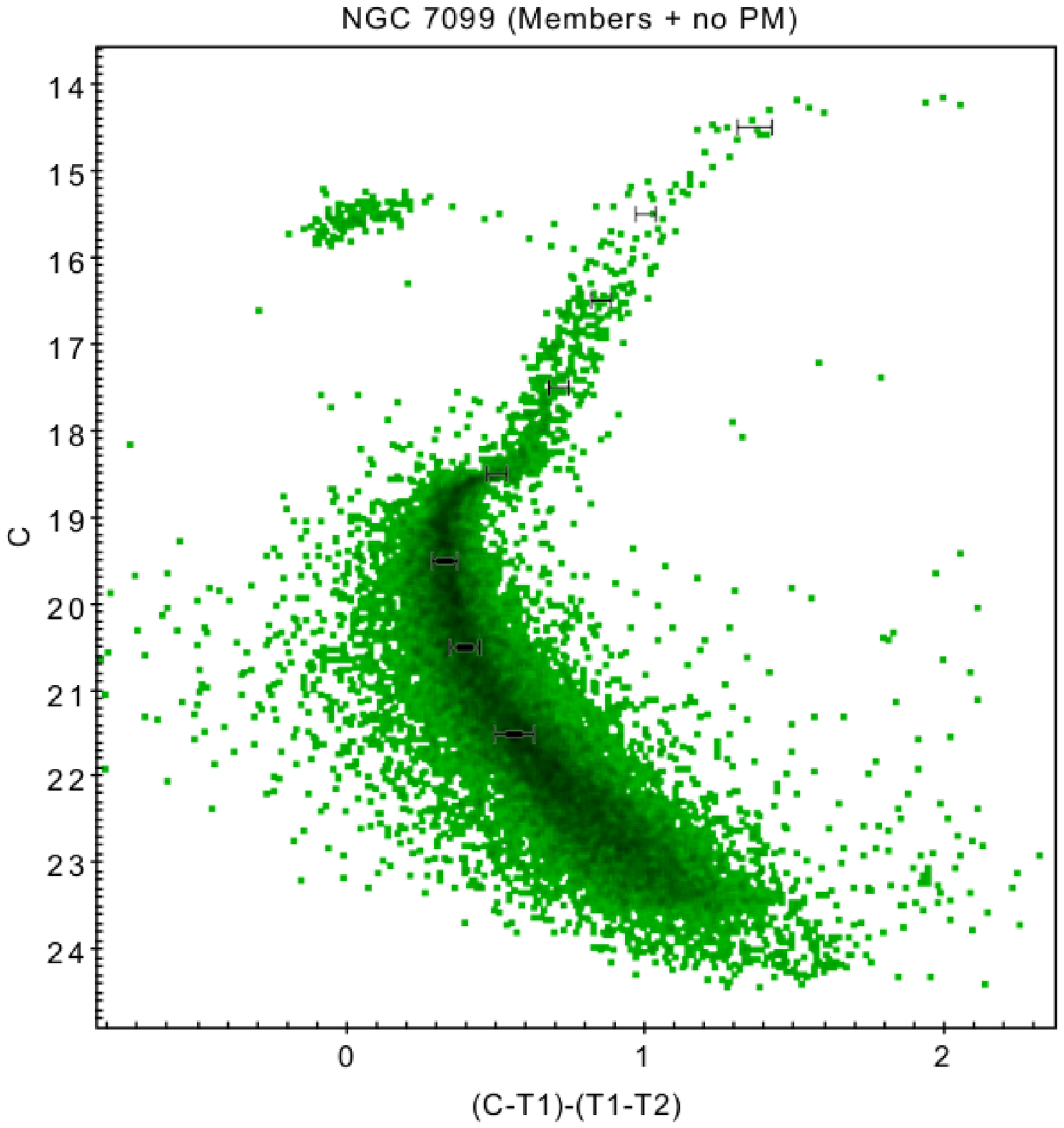}
	
    \caption[CMDs using the new color $(C-T1)-(T1-T2)$ vs $C$]{Left: The $(C-T1)$ CMDs from Figure \ref{gaia }. Right: The new color $(C-T1)-(T1-T2)$ vs $C$. Mean color and magnitude errorbars in 1 magnitude bins along the principal sequence are displayed as black crosses.
    }
    \label{CT1T2}
\end{figure*}

Until now, our best weapon in the Washington System to find MPs was to use the $C-T_1$ color and plot vs C in a CMD, leaving other C filter combinations to show partially defined MPs ($C-T_2$), or how the absence of the C filter fails to separate MPs ($T_1 - T_2$).
   But careful analysis of Figure \ref{synth} shows that the $T_2$ filter is roughly centered on multiple CN bands which have a fairly significant flux difference between 1P and 2P stars, allowing (in theory) to further separate the populations of the cluster, although our previous studies demonstrated that the spread in $T_1-T_2$ is almost completely due to the errors. Our hypothesis is that $T_2$  retains some capability to to distinguish MPs due to the CN-bands that it includes (as seen in figure \ref{synth}), but the separation of the different sequences is difficult to detect. So based on the technique from \citet{Piotto2015}, we created a new combination of colors: $(C-T_1)-(T_1-T_2)$ (or $C+T_2-2T_1$). The idea is that we can potentially further separate the sequences in a CMD combining the potential of $C-T1$ with a small additional difference generated in $T_1-T_2$. We also note that the C filter includes both CN bands as well as the CH band. The former are stronger in 2P vs. 1P stars due to the fact that the CN-band strength is controlled by the N abundance, which is enhanced in 2P over 1P stars. However, the CH band is weaker in 2P vs. 1P stars since C is depleted. Hence, these 2 effects work against each other to some extent, although it is also clear that the strongest effect is due to the various CN bands, so that the flux in the C filter will be less in a 2P star compared to that of an otherwise identical 1P star. Similarly, it should also be less in the T2 filter for a 2P vs 1P star.
  Figure \ref{CT1T2} shows the new $(C-T_1)-(T_1-T_2)$(hereafter $C,T_1,T_2$) vs $C$ CMDs.
  
  A detailed analysis is shown in the next subsections.

\subsection{NGC 7099}
\begin{figure}
\centering
  	\includegraphics[width=0.8\columnwidth]{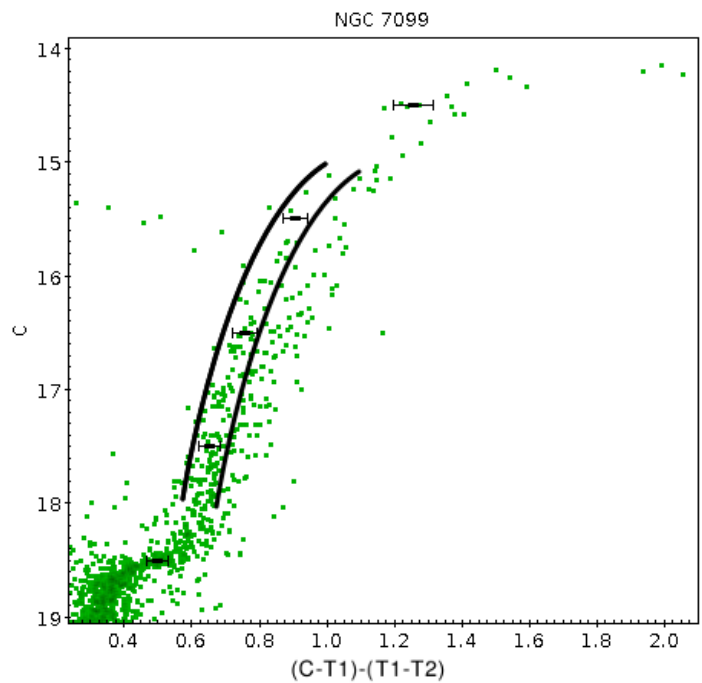}
    \caption{Criteria to divide the 1P from the 2P. The black lines represent the limits established for being the 1P.}
    \label{overlap}
\end{figure}

For NGC 7099, in both CMDs (pm members and members + stars with no pm), we can see a very broad RGB compared to $C-T_1$, similar to the CMD shown in \citet{Piotto2015} for the same cluster.
Based on the $C,T_1,T_2$ CMDs of NGC 7099 from figure \ref{CT1T2} we included the mean color error in 1 magnitude bins along the principal sequence, but attached to the left border of the RGB. We consider these stars inside the errorbars as those associated with the 1P, while those at the right of the limits of the errorbars are considered as 2P stars, so taking this as a guide we established our 1P/2P division in the catalog containing members + stars with no PM by drawing two lines, each connecting with the limits of the error bars, at both sides, as seen in Figure \ref{overlap}. The samples of each population of stars is taken between the 15-18 magnitude range in C and 13-17 in T2 since the MPs begin to merge in the brighter bins, and the AGB complicates the separation as well.

In this part we made a correction in F17. There, we took a group of stars at the left of the RGB deemed as the 1P. Comparing the radial distributions of both 1P and 2P stars of NGC 7099 we got the most impressive but unexpected conclusion of the publication: The first population of NGC 7099 was more centrally concentrated than the second, opposite to most of the actual observations as well as MP formation models. However, our new research proves that conclusion to be wrong, since the 1P subset of F17 does not appear in our new CMD, meaning that probably it was composed of field stars that could not be removed then given the absence of Gaia PMs at the time. This would also explain why we got a p-value of 0 in our Kolmogorov-Smirnov test, \footnote{If P<0.05, one must reject the null hypothesis of no difference between two datasets, more information about this test is found in http://www.physics.csbsju.edu/stats/KS-test.html} indicating that the 1P and 2P subsets were different distributions.

We now take a new subset in $C-T_1$, trying to replicate the one from $C,T_1,T_2$ assumed to be our 2P and leaving the rest of the RGB as the 1P. These subsets were compared in the other colors and their radial distributions were tested to analyze which pair of subsets was more effective in distinguishing the MPs. 
What we should expect in this part is to have the 1P at the blue side and the 2P at the red side in both $C-T_1$ and $C,T_1,T_2$ colors with, hopefully, better defined subsets in the latter, but with totally the opposite ocurring in T1-T2, since in this color the filter that covers the CN-band appear as the subtrahend (and this explains why subtracting this color from $C-T_1$ should help to increase the spread on the RGB).

\begin{figure*}
  \includegraphics[width=0.32\linewidth]{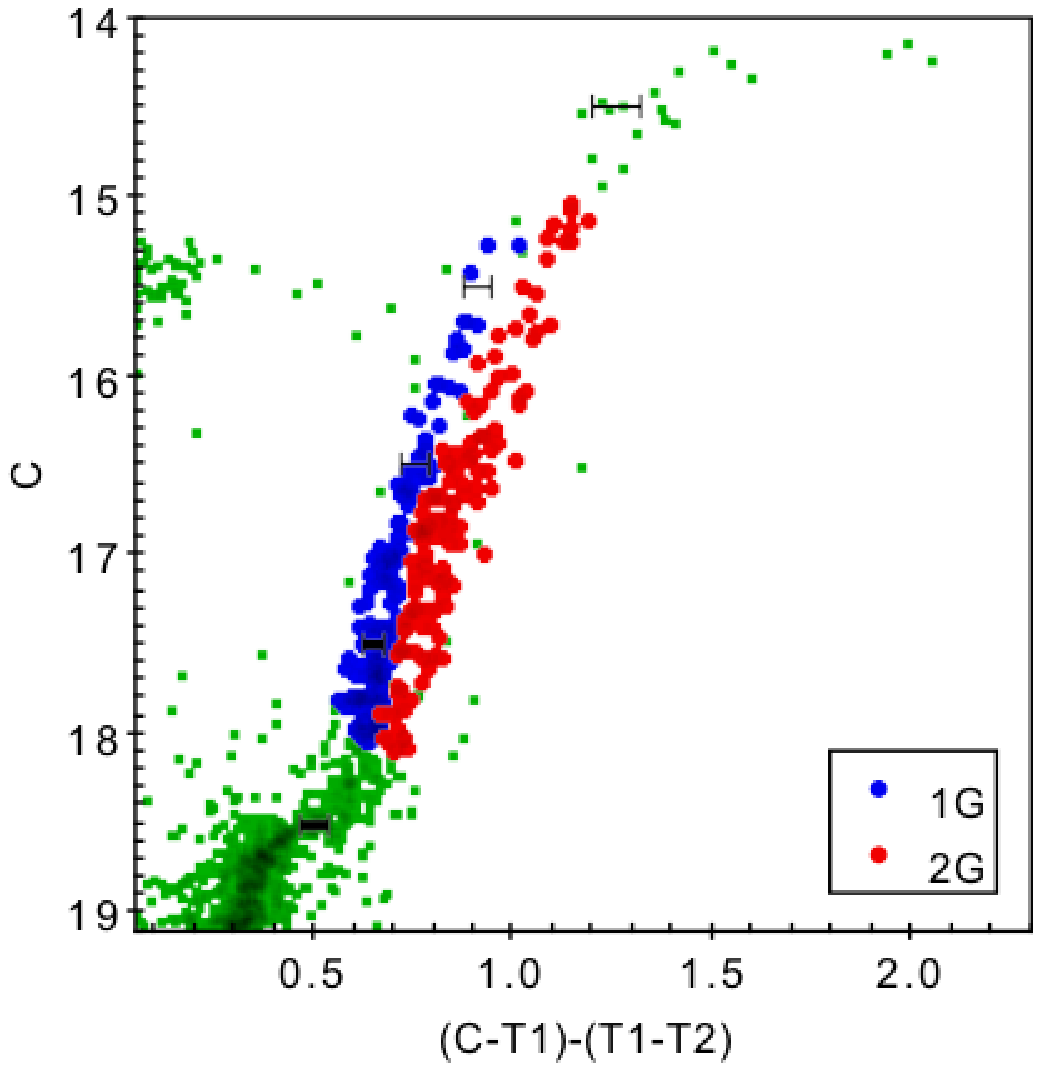}
  	\includegraphics[width=0.32\linewidth]{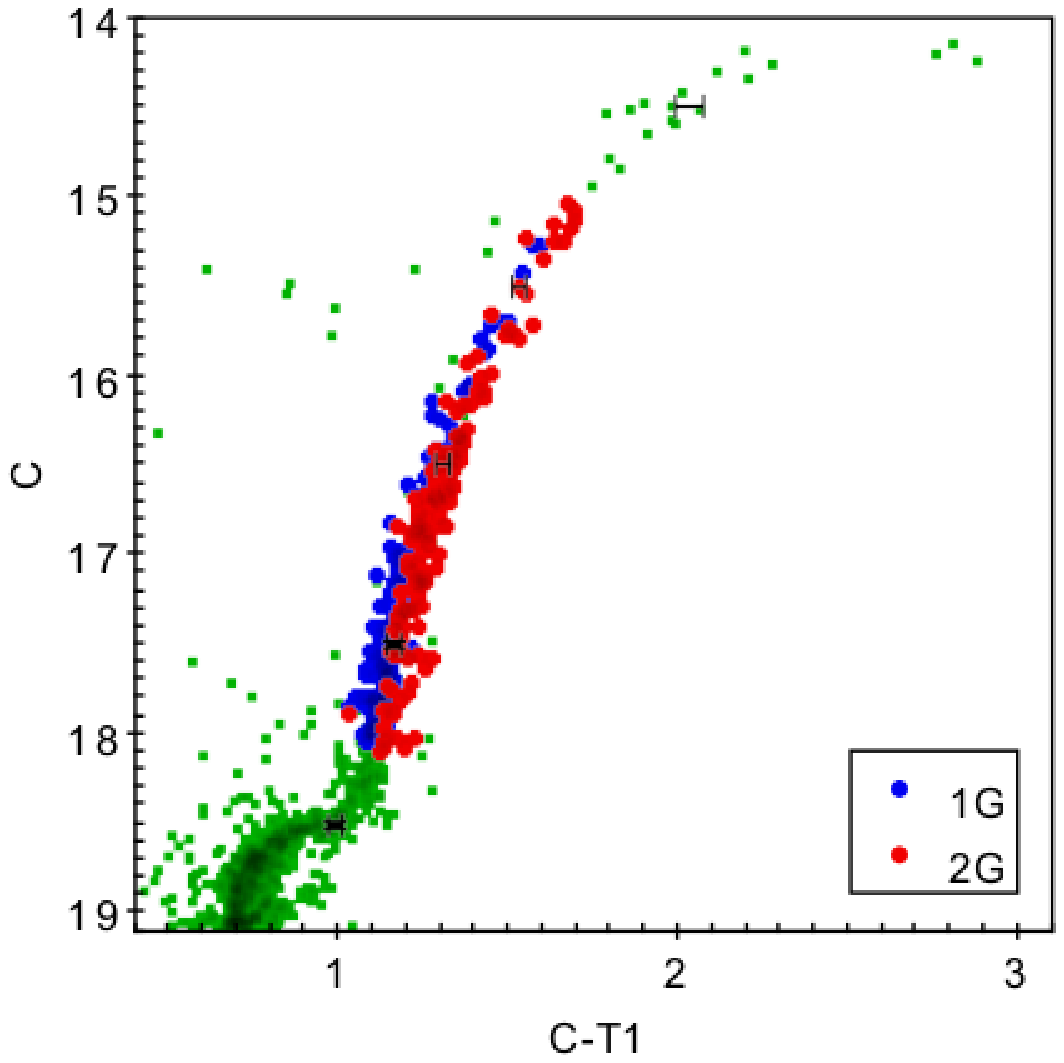}
  	\includegraphics[width=0.32\linewidth]{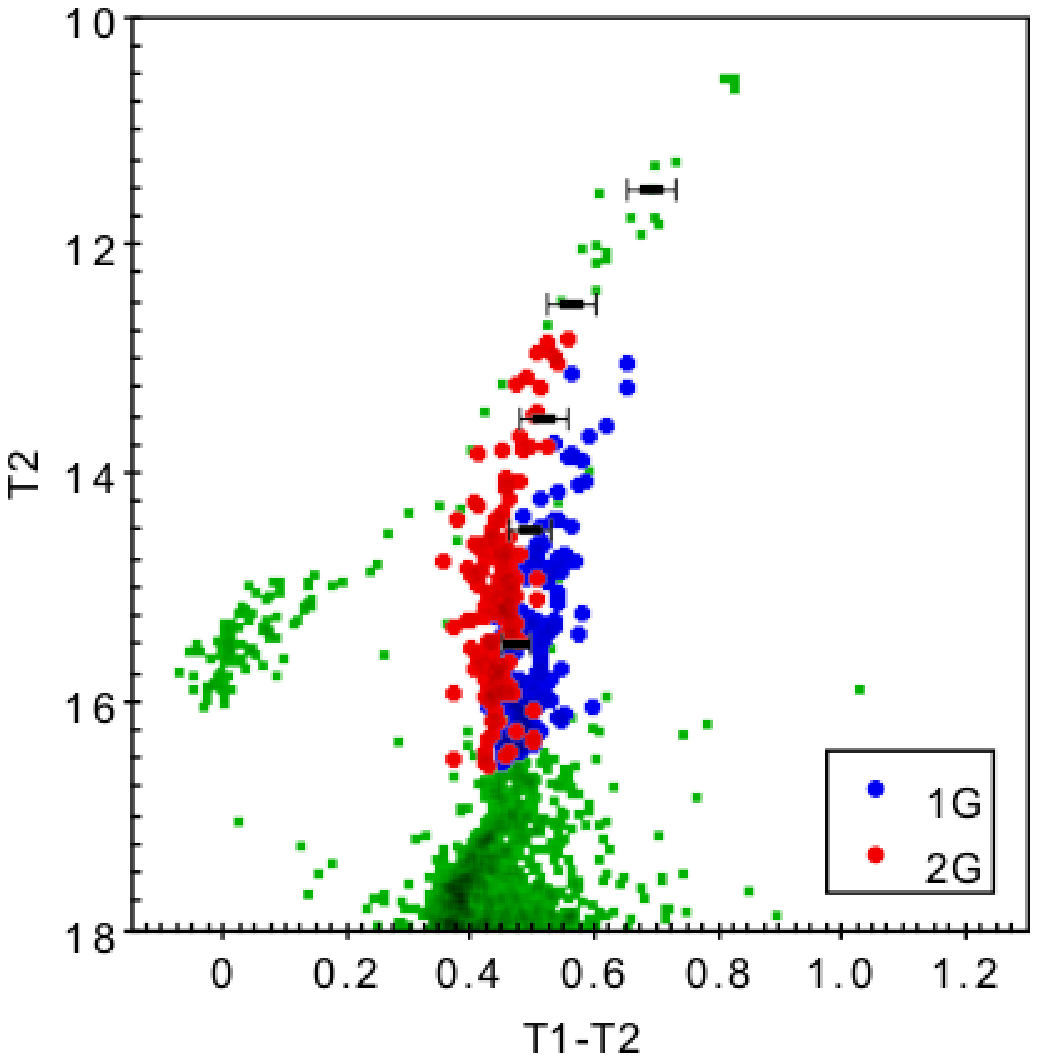}
  	
  	\includegraphics[width=0.32\linewidth]{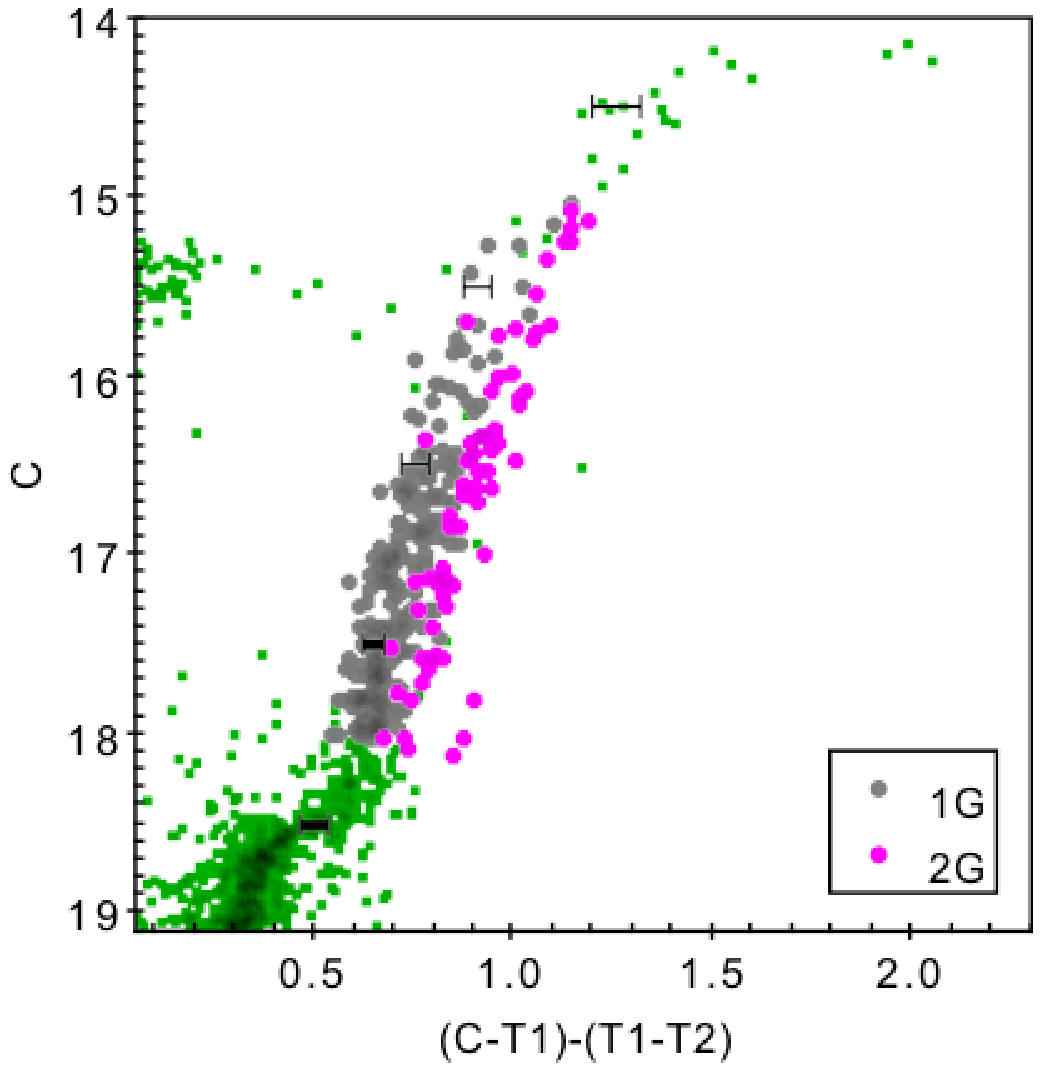}
  	\includegraphics[width=0.32\linewidth]{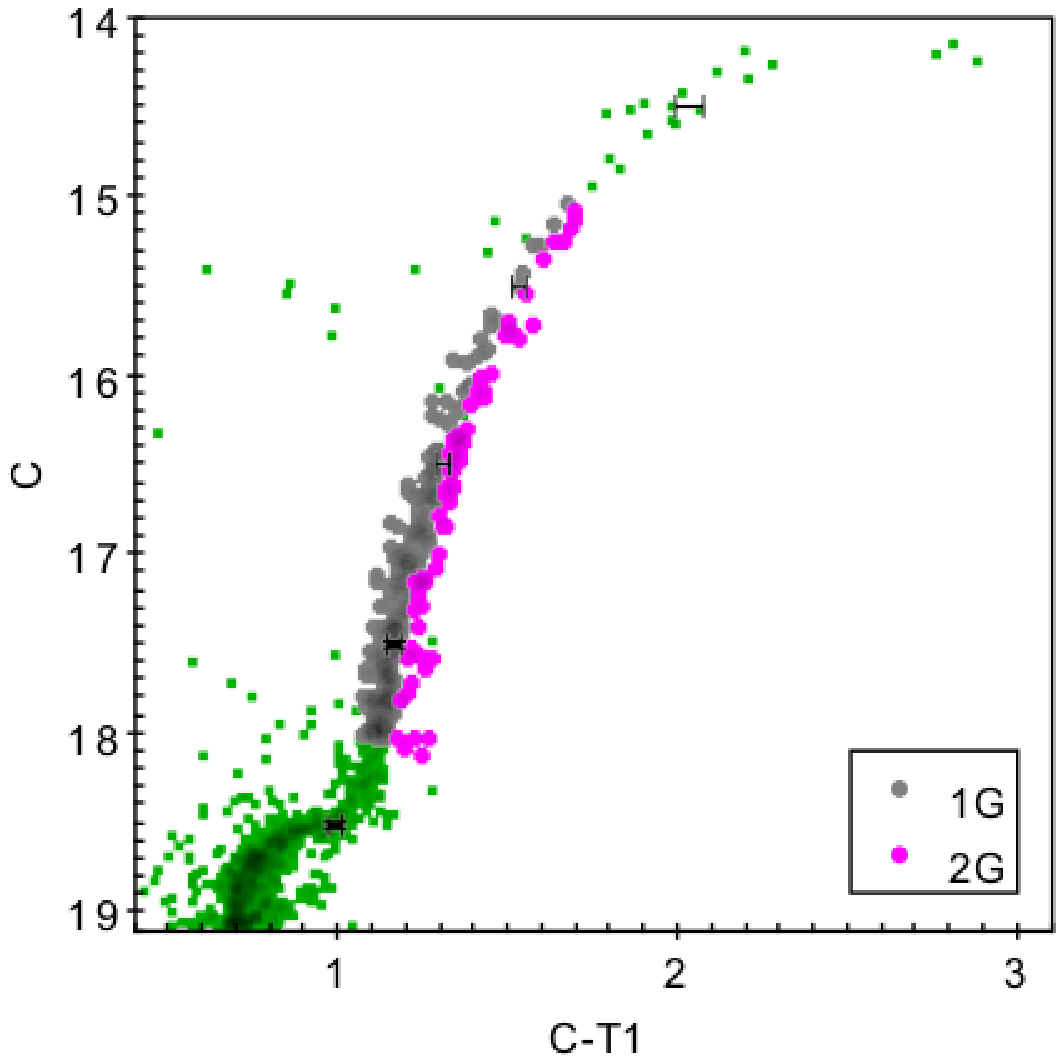}
  	\includegraphics[width=0.32\linewidth]{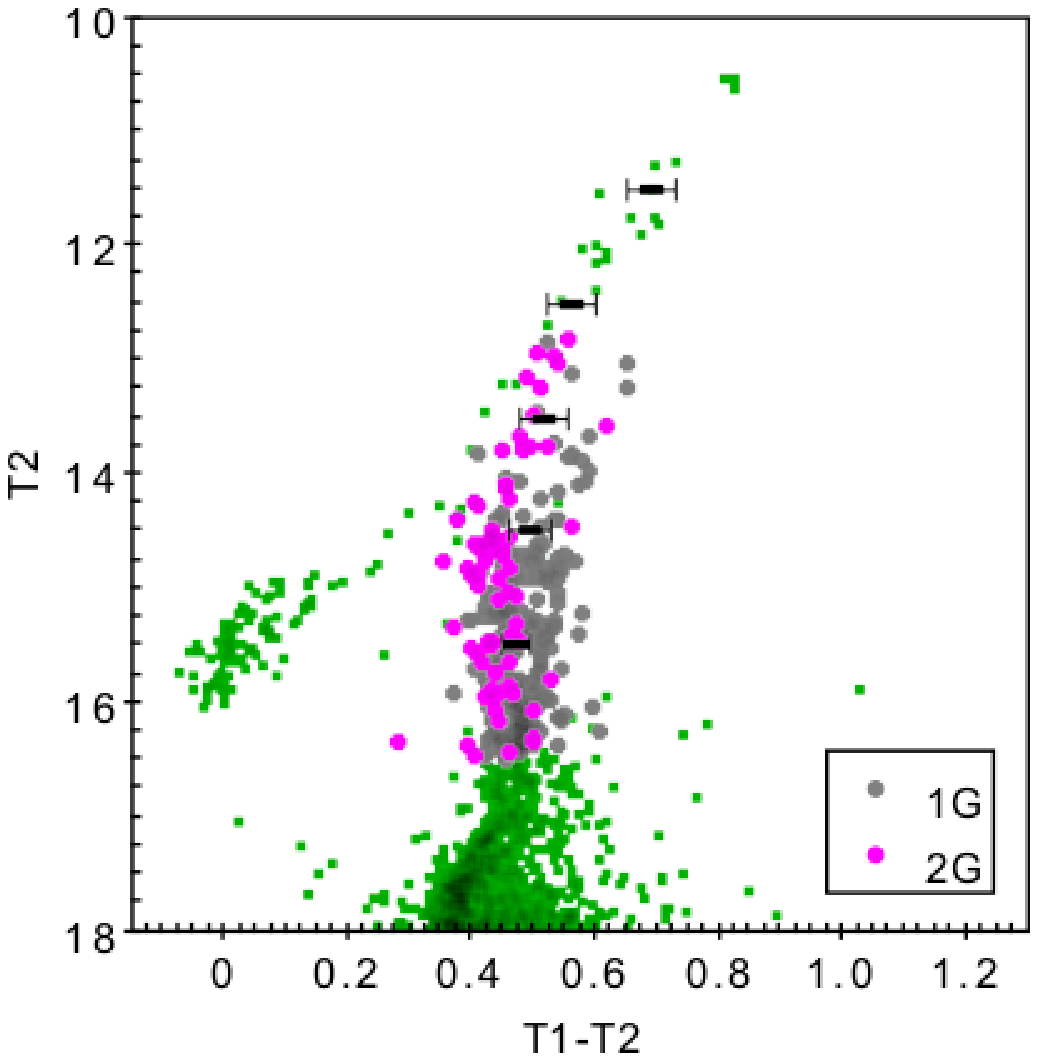}

  	\includegraphics[width=0.32\linewidth]{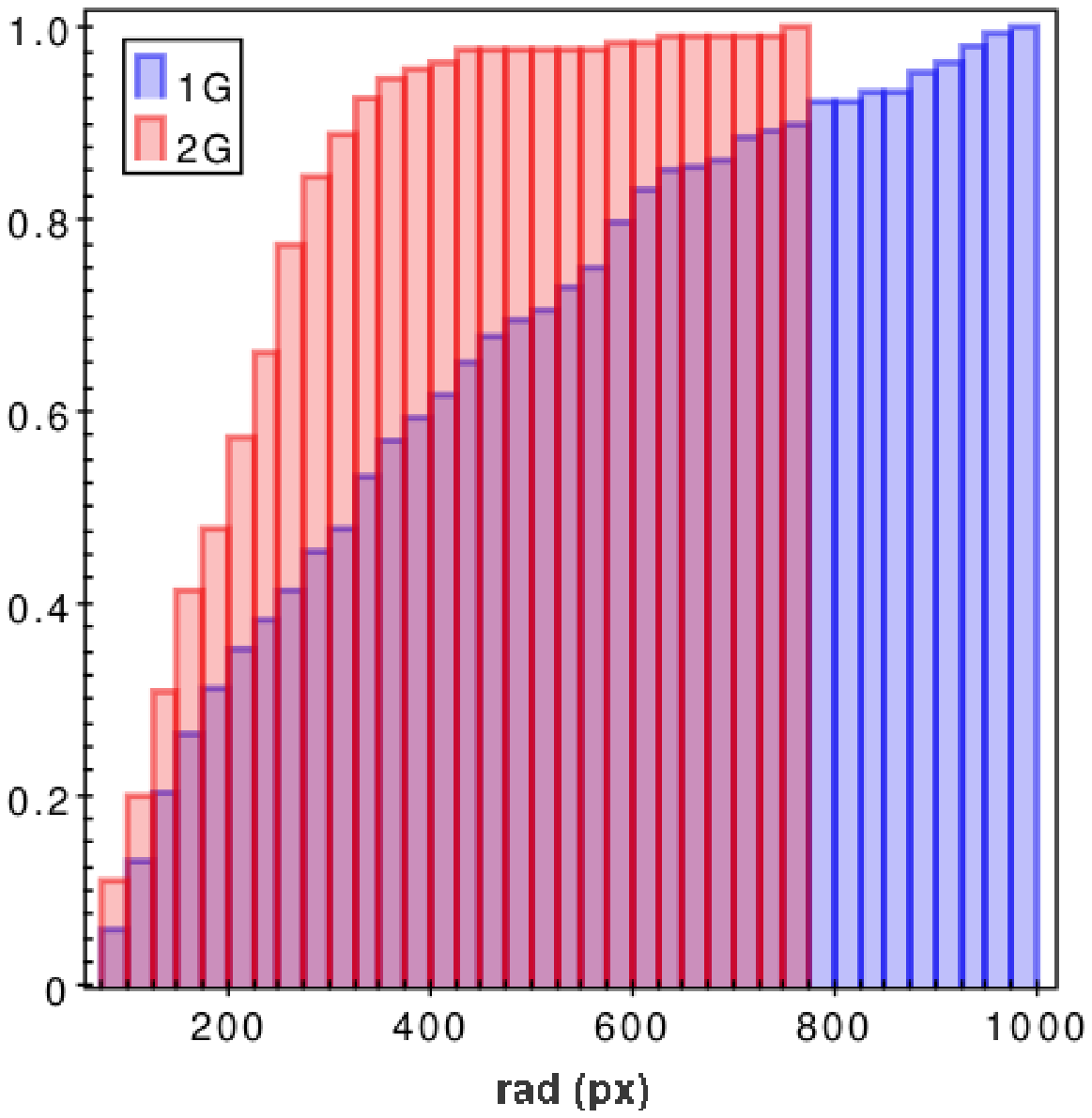}
  	\includegraphics[width=0.32\linewidth]{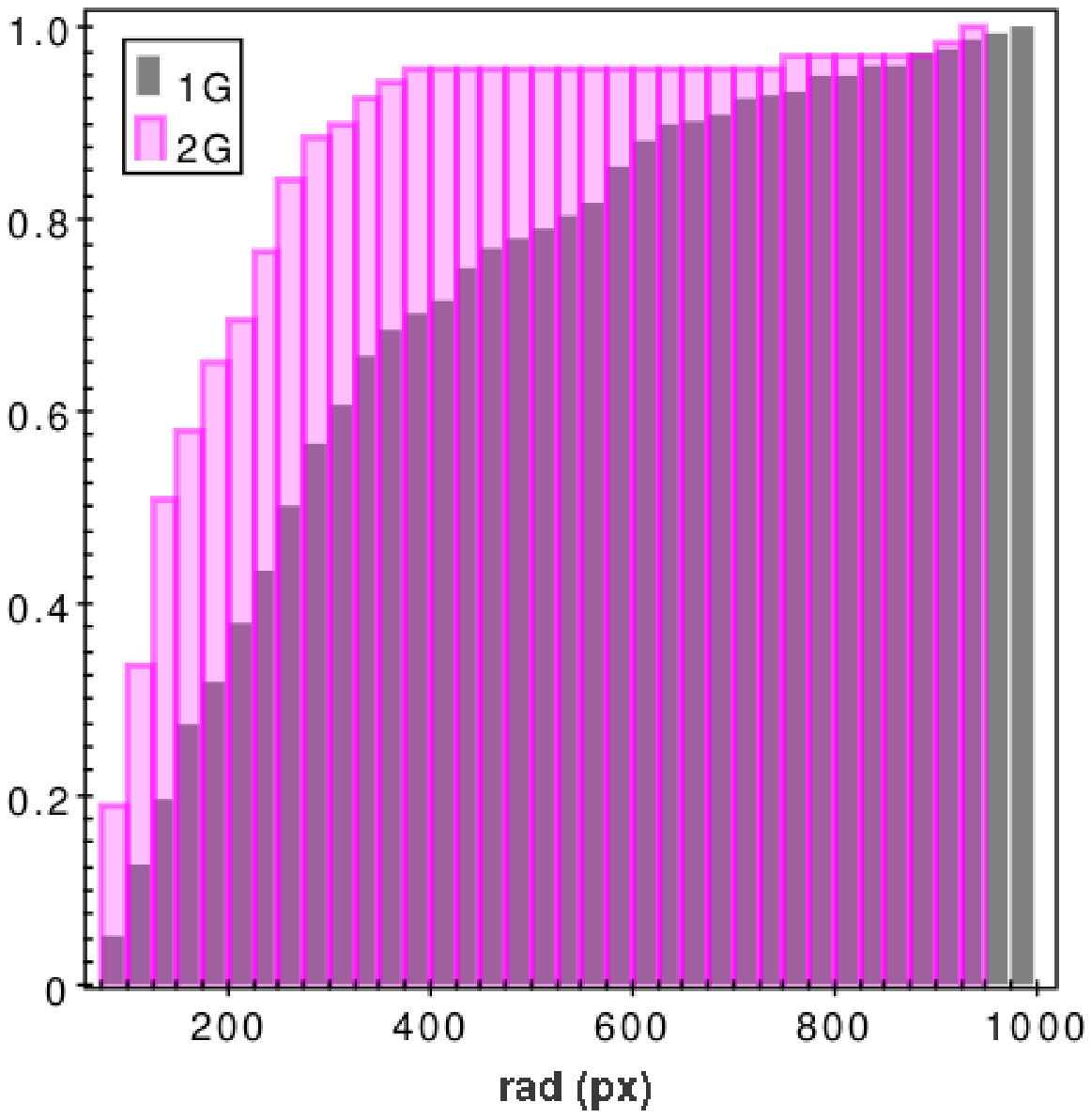}
  	\includegraphics[width=0.32\linewidth]{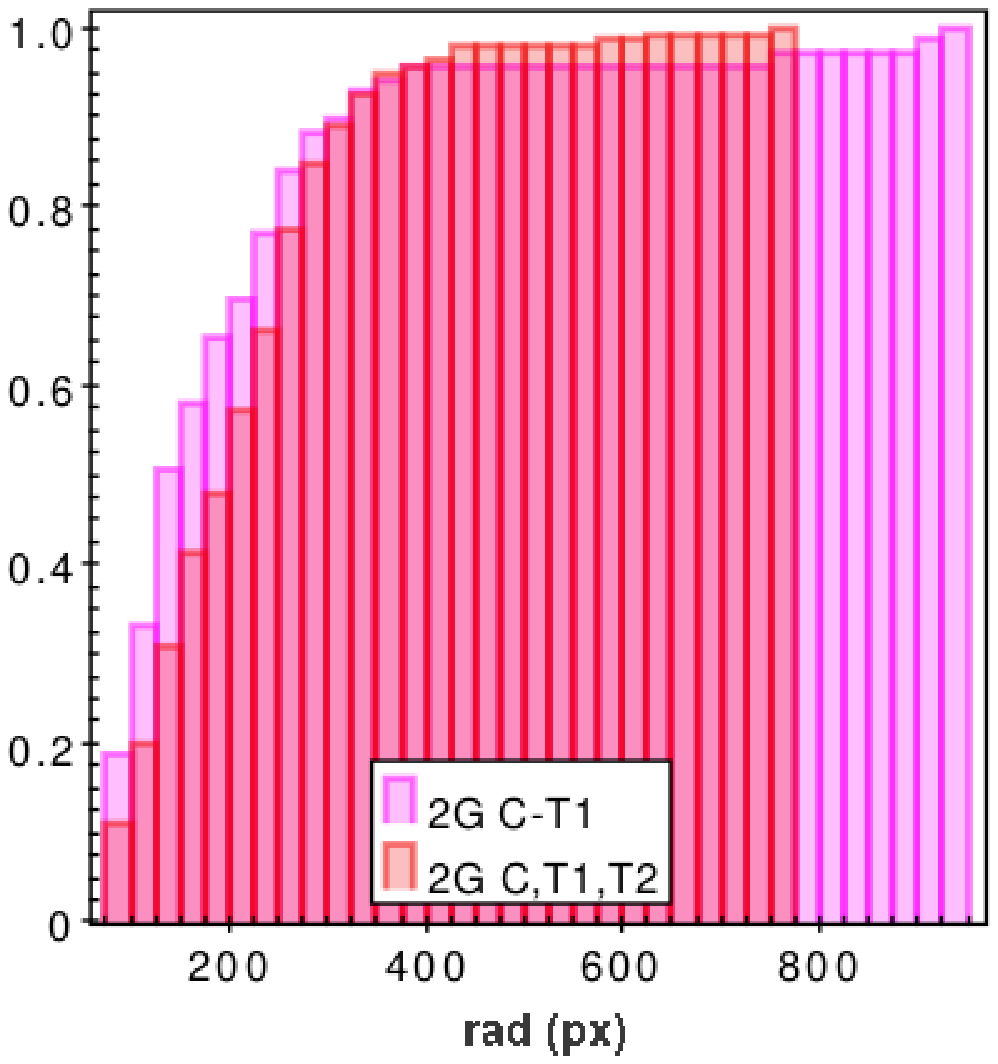}
  	
    \caption{NGC 7099: Upper panels: A 1P/2P subset chosen from the color C,T1,T2 shown in C-T1 and T1-T2.
    Mid panels: A 1P/2P subset chosen from the color C-T1 shown in C,T1,T2 and T1-T2.
    Bottom panels: Left: Radial distributions of 1P and 2P of the subset from C,T1,T2. Middle: Radial distributions of 1P and 2P of the subset from C-T1. Right: Comparison of the 2P from the C,T1,T2 and the 2P from C-T1.}
    \label{Comp7099}
\end{figure*}

The results are shown in Figure \ref{Comp7099}. As expected, the subsets made based on the color $C,T_1,T_2$ (upper panels) are a bit less defined in the CMD with $C-T_1$, since the separation in  the latter color should be less than that in the former. In $T_1-T_2$ both populations seem to be well separated but mirrored.

The subsets made from $C-T_1$ (middle pannels) show a less effective separation in $C,T_1,T_2$ and a very similar one in $T_1-T_2$. Both groups of CMDs look very similar. With $C-T_1$ the percentage of 2P stars is 23.2\%$\pm$25\% of the RGB while for $C,T_1,T_2$ the percentage of 2P is 44.9\%$\pm$16,7\%. This big difference in percentages is due to the smaller spread in $C-T_1$, causing that any small error in selecting a subset includes/excludes lots of stars. This is seen in the high percentage of error in $C-T_1$ (higher even than the percentage of the population itself) obtained putting the errorbars in the middle of our 1P/2P division and enclosing all the stars inside their limits to see whose stars that could be being included/excluded in our 1P/2P selection respect to the total number of stars in the RGB.

Comparing the radial distributions of both pairs of 1P/2P stars (lower pannels) we can see that both of them show a 2P more centrally concentrated, in agreement with most of the MP formation scenarios. In fact, both pair of subsets give a P-value = 0 in a K-S test. While the subset from $C-T_1$ grows faster with radius than $C,T_1,T_2$, the latter color includes all stars after reaching $\sim$780px from the center($\sim$200px less than $C-T_1$). The lowest right panel compares both 2P groups of stars. A K-S test between these give a P-value of 0.026, indicating that we should reject the null hypothesis of no difference between both distributions, so there are significant differences while selecting a subset from $C-T_1$ or $C,T_1,T_2$. And in fact, since C-T1 shows the strongest central concentration in the inner parts, its behaviour is preferred in this regard.

\begin{table}
\centering
\caption{Comparison of error vs width in NGC 7099.} \label{7099vs}
\begin{tabular}{llll}\toprule
Magnitude Range & Mean width & Mean error & Ratio\\\midrule
& C-$T_1$ vs C\\\midrule
15-16 & 0.043 & 0.021 & 2.05 \\   
16-17 & 0.038 & 0.016 & 2.38 \\
17-17.9 & 0.041 & 0.016 & 2.56 \\\midrule
& $T_1$-$T_2$ vs $T_2$\\\midrule
13-14 & 0.061 & 0.039 & 1.56\\
14-15 & 0.052 & 0.033 & 1.58\\
15-16.4 & 0.042 & 0.022 & 1.91\\\midrule
& C,$T_1$,$T_2$ vs C \\\midrule
15-16 & 0.073 & 0.043 & 1.70 \\
16-17 & 0.070 & 0.035 & 2.00 \\
17-17.9 & 0.063 & 0.027 & 2.33 \\\bottomrule

\end{tabular}

\end{table} 

Table \ref{7099vs} shows the standard deviation (read as the width) of the RGB and the mean error in bins of 1 magnitude (with the exception of the last bins), for each of the 3 colors, while the last column shows the ratio width/error, which is a robust measurement of the effectiveness of the colors for separating MPs.
 As expected, $T_1-T_2$ shows a mean ratio of 1.68, in agreement with previous studies that indicates that, although this color shows a spread slightly bigger than the errors, it is not very sensitive to the presence of MPs. The small observed difference could in fact come from an undetected source of error. For $C-T_1$, the mean ratio is 2.33, significantly larger, and enough to assure the detection of MPs. Unfortunately, $C,T_1,T_2$ only has a mean ratio of 2.01, significantly less than that of $C-T1$. However, we note that this is not unexpected since it is not a combination of new filters but only a combination of the width and error of the first 2 colors.

\subsection{NGC 1851}

\begin{figure*}
\centering
  \includegraphics[width=0.32\linewidth]{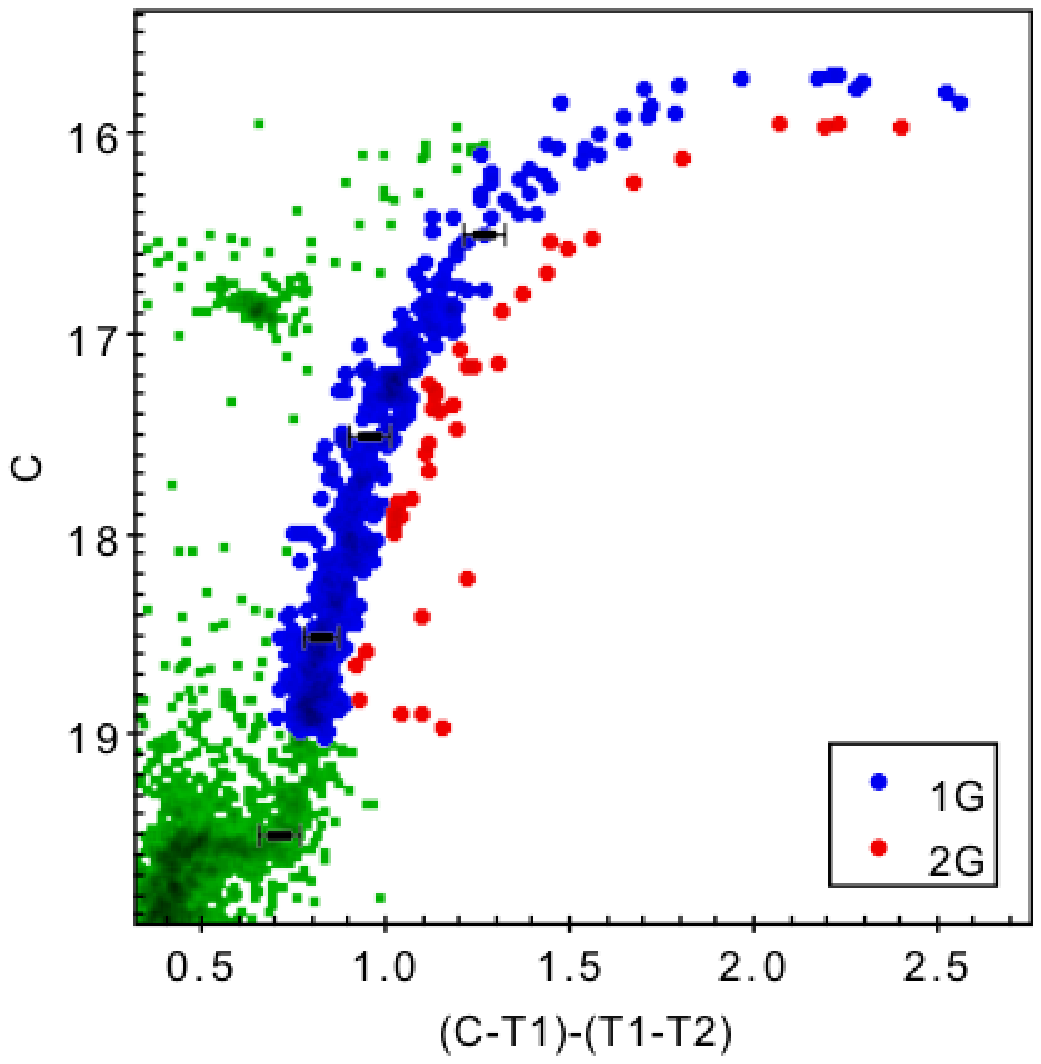}
  \includegraphics[width=0.32\linewidth]{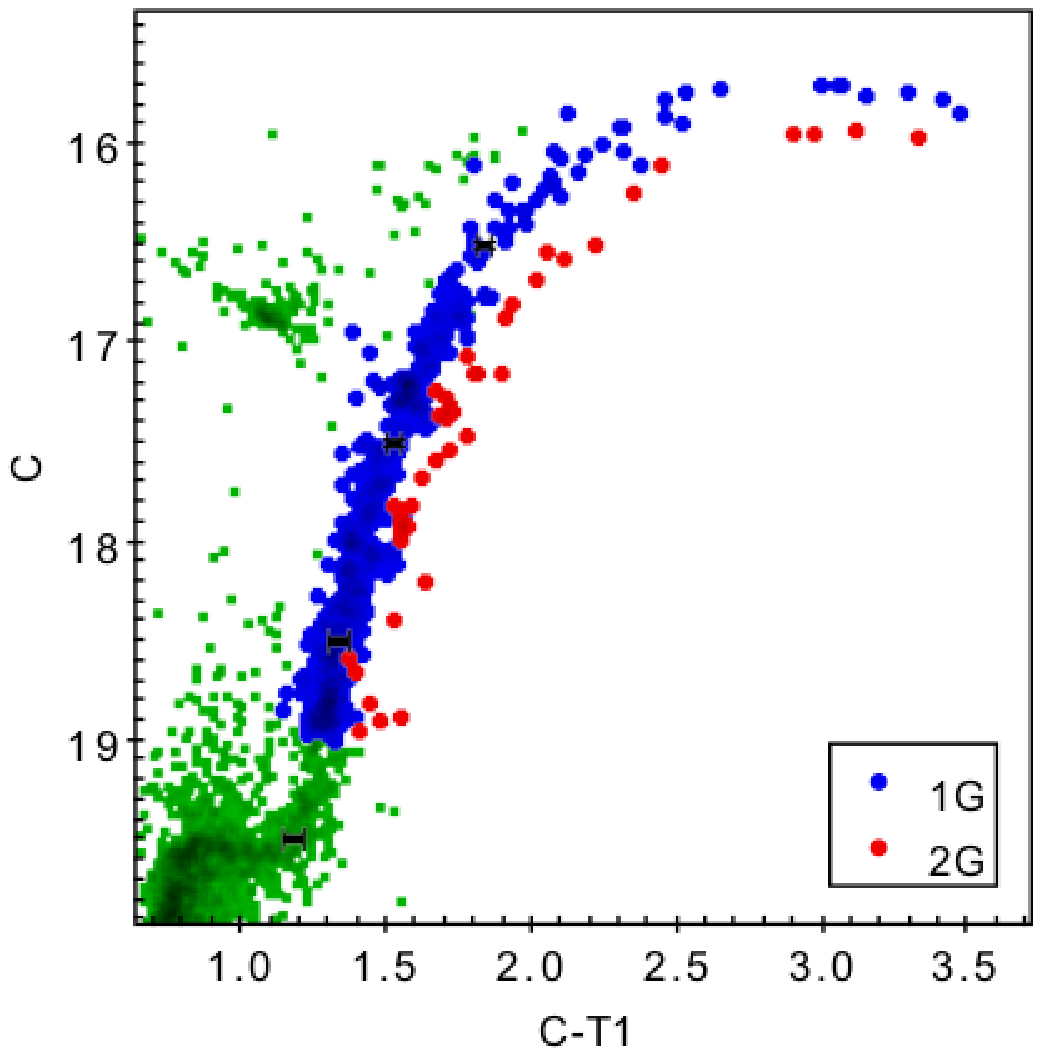}
  \includegraphics[width=0.32\linewidth]{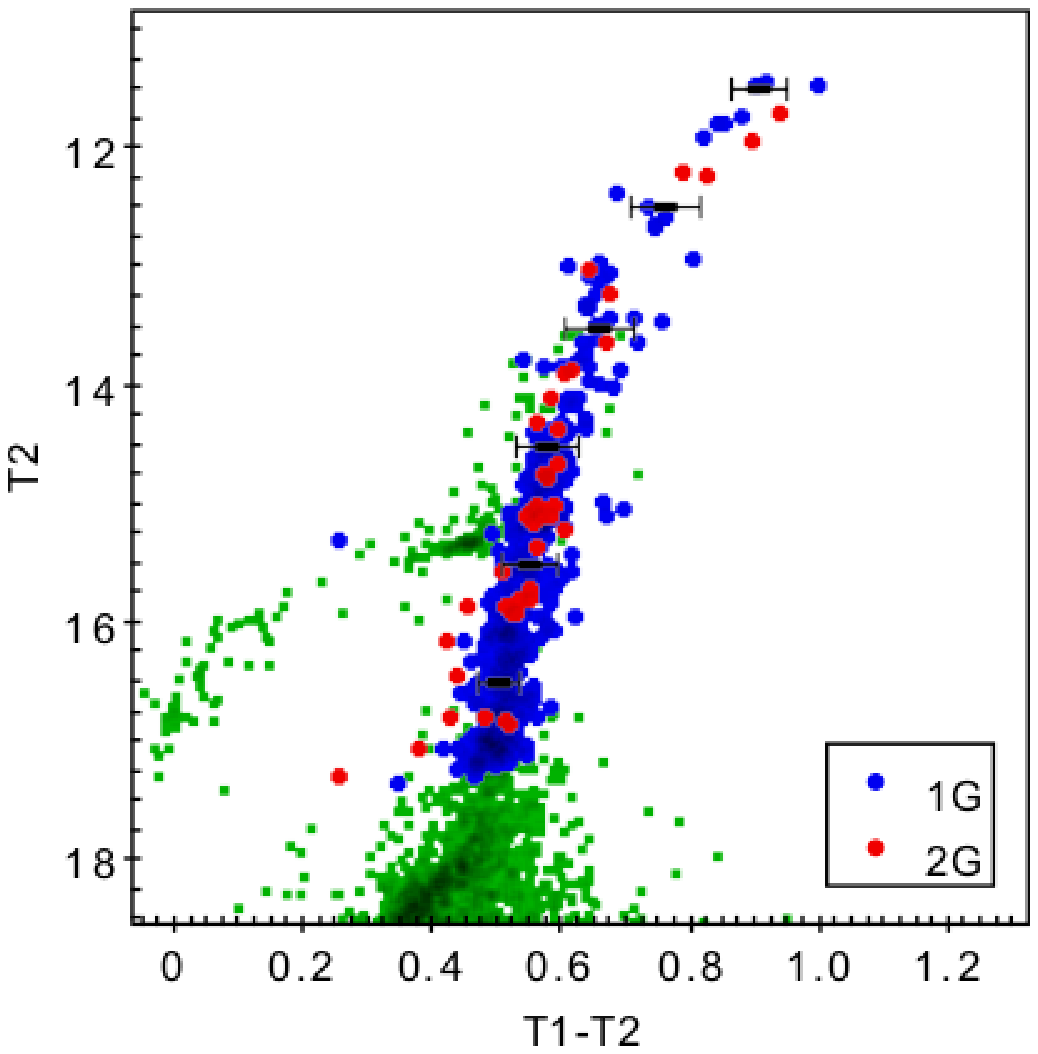}
  	
  \includegraphics[width=0.32\linewidth]{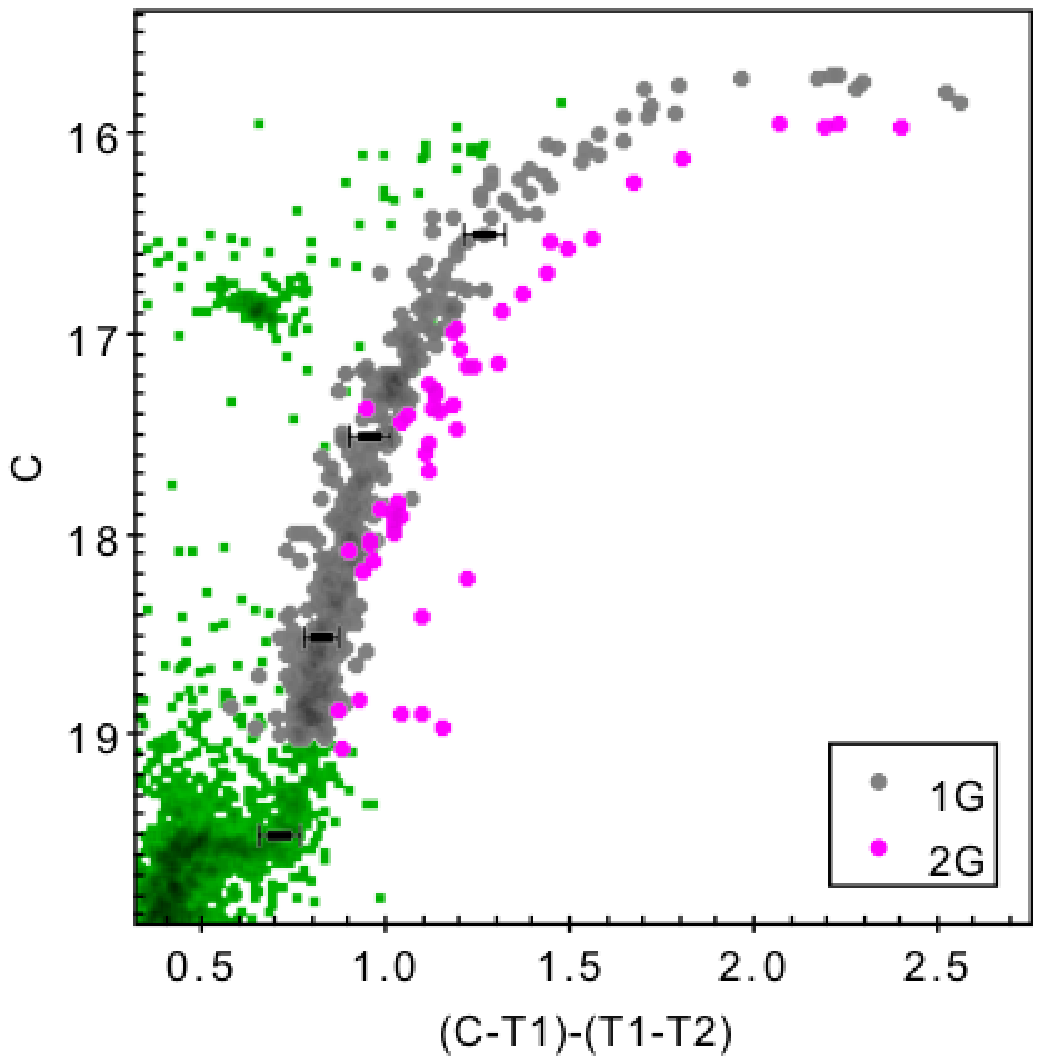}
  \includegraphics[width=0.32\linewidth]{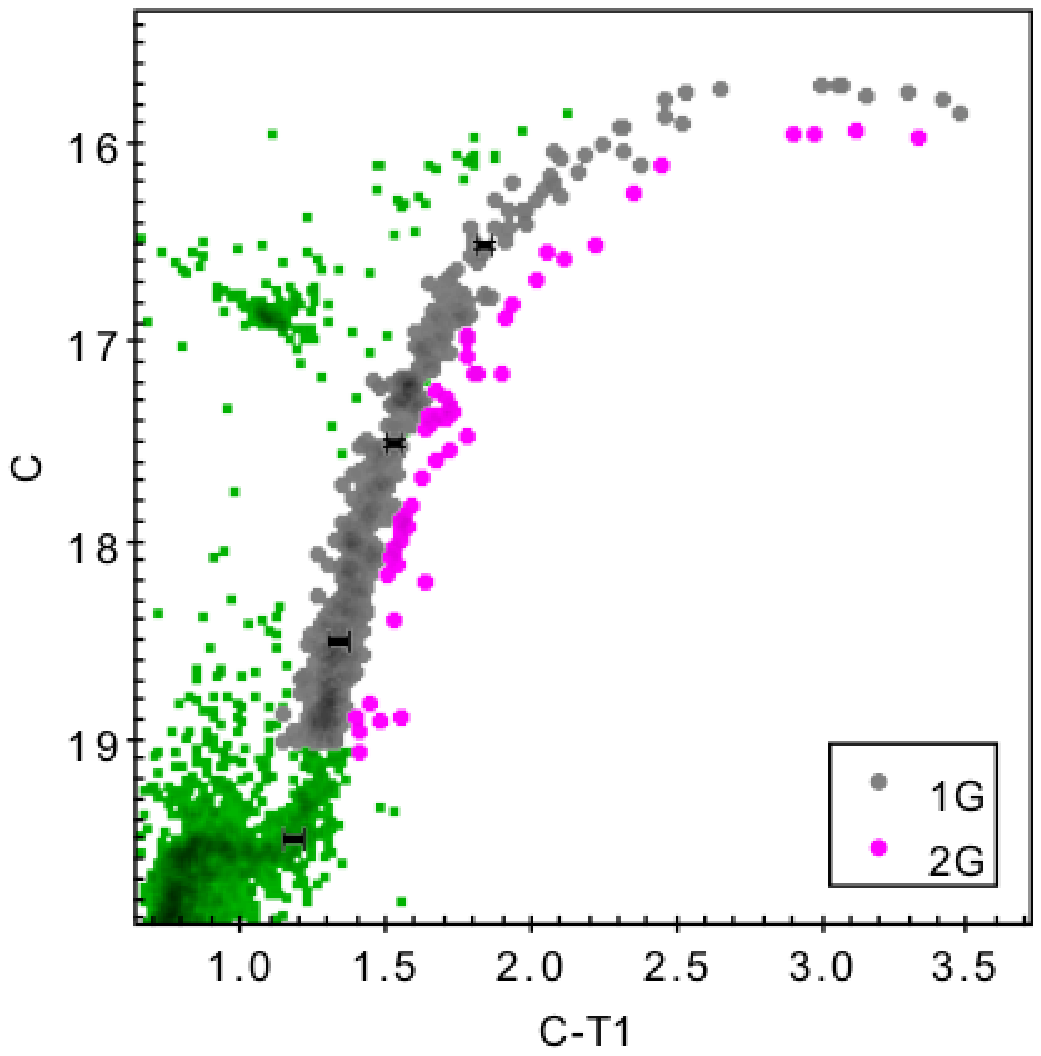}
  \includegraphics[width=0.32\linewidth]{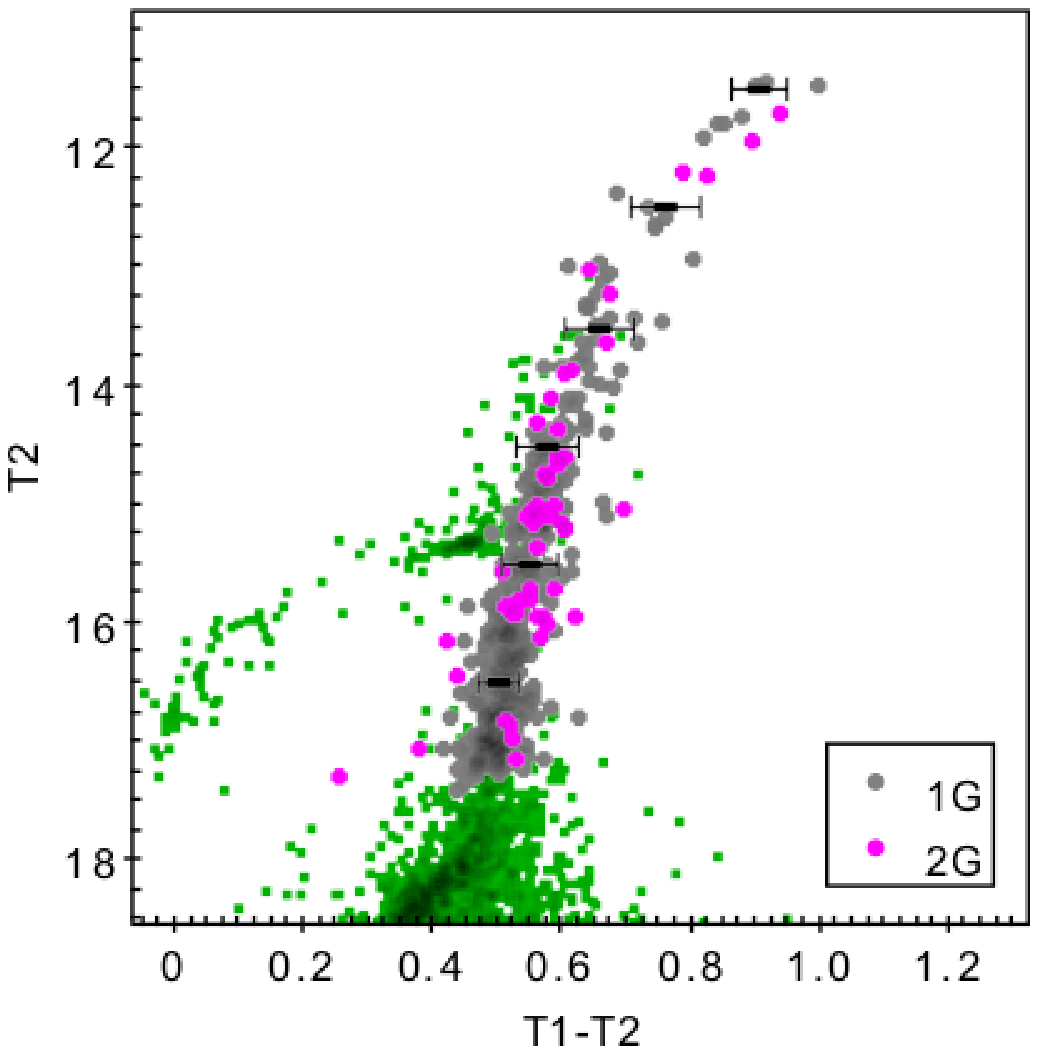}
  	
  \includegraphics[width=0.32\linewidth]{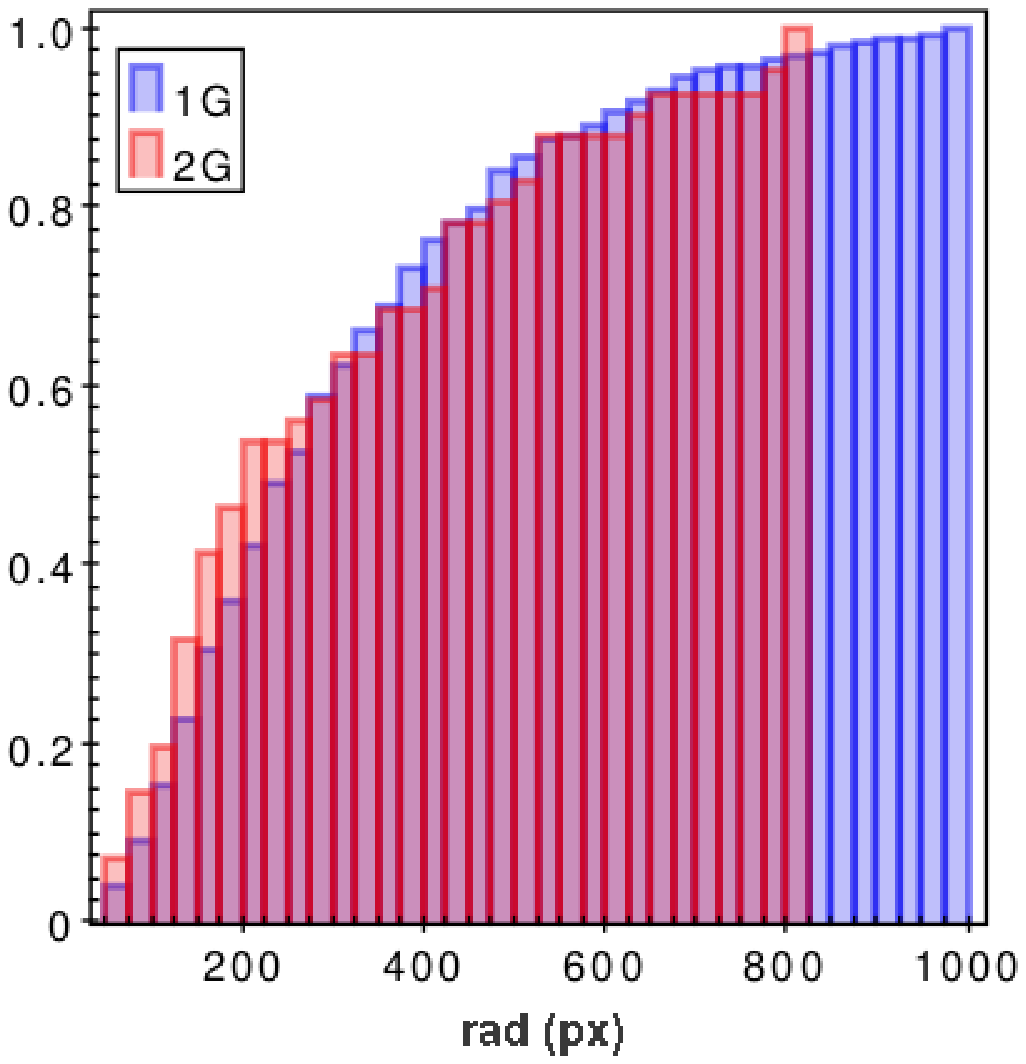}
  \includegraphics[width=0.32\linewidth]{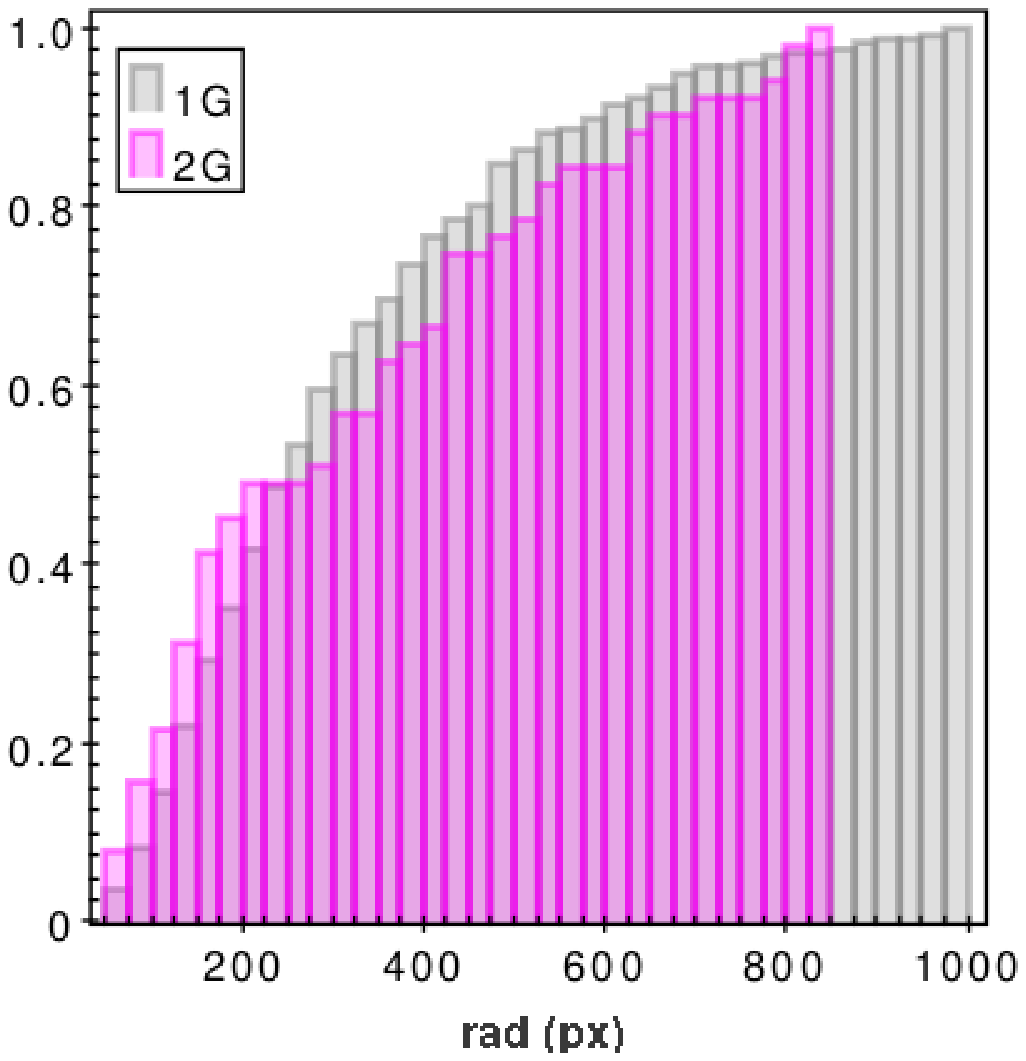}
  \includegraphics[width=0.32\linewidth]{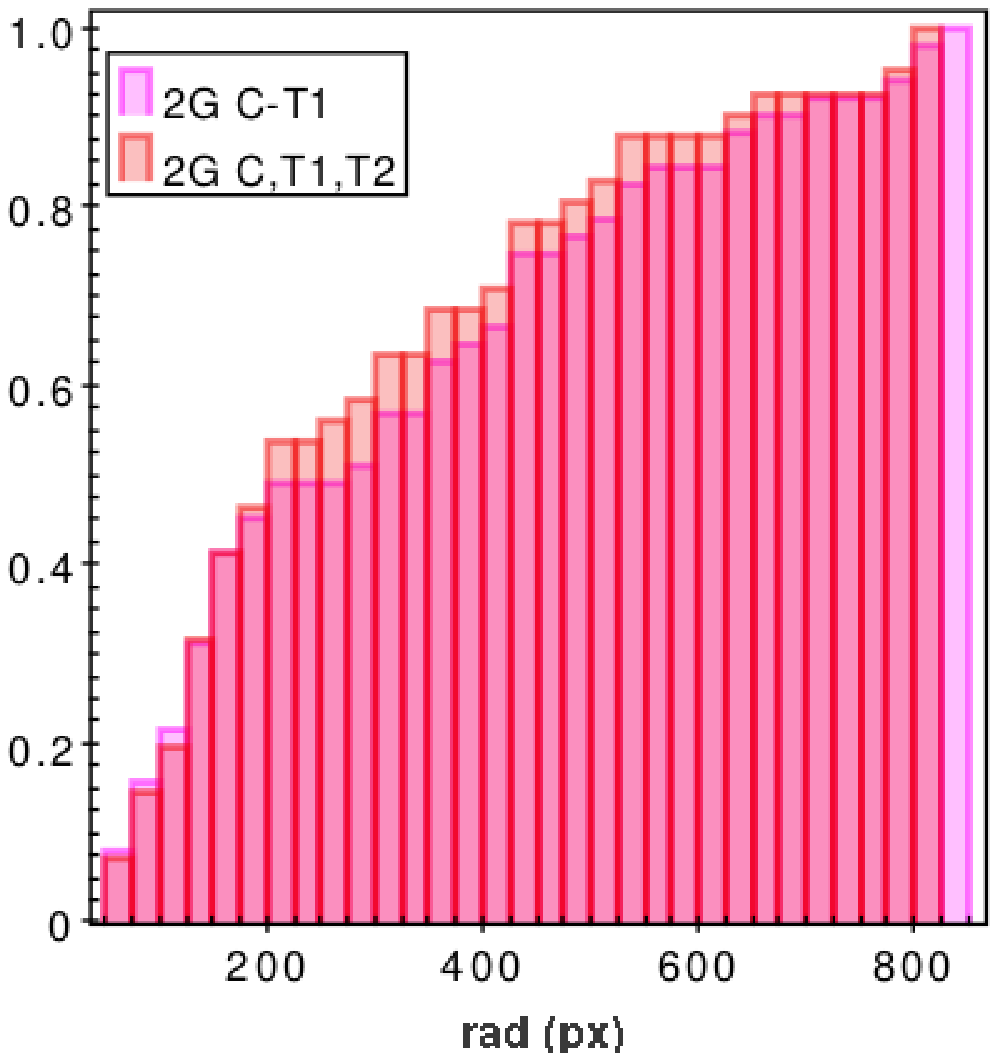}

    \caption{NGC 1851: Upper panels: A 1P/2P subset chosen from the color C,T1,T2 shown in C-T1 and T1-T2.
    Mid panels: A 1P/2P subset chosen from the color C-T1 shown in C,T1,T2 and T1-T2.
    Bottom panels: Left: Radial distributions of 1P and 2P of the subset from C,T1,T2. Middle: Radial distributions of 1P and 2P of the subset from C-T1. Right: Comparison of the 2P from the C,T1,T2 and the 2P from C-T1.}
    \label{Comp1851}
\end{figure*}

\emph{-The Blue and Red RGBs}\\
As shown in Figure \ref{Comp1851}, NGC 1851 presents a double RGB: The left RGB, hereafter the Blue-RGB; and the right RGB, hereafter the Red-RGB. Both sequences are already well divided in $C-T_1$, so there are no clear differences in the subsets made from $C-T_1$ or $C,T_1,T_2$. Indeed, the amount of Red-RGB stars is 11.4\% and 9.3\% in $C-T_1$ and $C,T_1,T_2$, respectively. A K-S test done to the Red-RGB of both subsets give a P-value of 0.995, meaning that there is no significant difference between them.
Anyway, a K-S test in both Blue-RGB and Red-RGB populations of $C,T_1,T_2$ give a P-value of 0.423 while for $C-T_1$ is 0.313; again neither comparison shows a significant difference. Both subsets also show very similar radial distributions, but as explained in C14, this behaviour could be due to the small amount of Red-RGB stars in both subsets. Also, opposed to what we expected, the Red-RGB does not appear at the right of the Blue-RGB in $T_1-T_2$ but dispersed along the entire Blue-RGB.

\begin{table}
\centering
\caption{Comparison of error vs width in NGC 1851.}\label{1851vs}
\begin{tabular}{llll}\toprule
Magnitude Range & Mean width & Mean error & Ratio\\\midrule
& C-$T_1$ vs C\\\midrule
15.7-17 & 0.078 & 0.031 & 2.52 \\
17-18 & 0.069 & 0.025 & 2.76 \\   
18-19.2 & 0.054 & 0.025 & 2.16 \\\midrule
& $T_1$-$T_2$ vs $T_2$\\\midrule
11.4-13 & 0.068 & 0.051 & 1.33\\
13-14 & 0.038 & 0.051 & 0.75\\
14-15 & 0.032 & 0.046 & 0.70\\
15-16 & 0.030 & 0.045 & 0.67\\
16-17 & 0.025 & 0.032 & 0.78\\
17-17.5 & 0.022 & 0.032 & 0.69\\\midrule
& C,$T_1$,$T_2$ vs C \\\midrule
15.7-17 & 0.074 & 0.056 & 1.32\\
17-18 & 0.068 & 0.049 & 1.39\\
18-19.2 & 0.060 & 0.041 & 1.46\\\bottomrule

\end{tabular}
\end{table} 

The ratio width/error in NGC 1851 (table \ref{1851vs}) shows that, although $C-T1$ give ratios even better than in NGC7099, the ratios in $T_1-T_2$ are very low, indicating a spread completely due to the errors. Naturally, the ratios of $C,T_1,T_2$ should be between the values of $C-T_1$ and $T_1-T_2$, as they are. And again, opposite to our original hope, $C,T_1,T_2$ does not show an improvement in the ratio with respect to $C-T_1$ and in fact is substantially worse, indicating that $T1-T2$ is not collaborating to help split the sequences.

\begin{figure}
  	\includegraphics[width=0.495\textwidth]{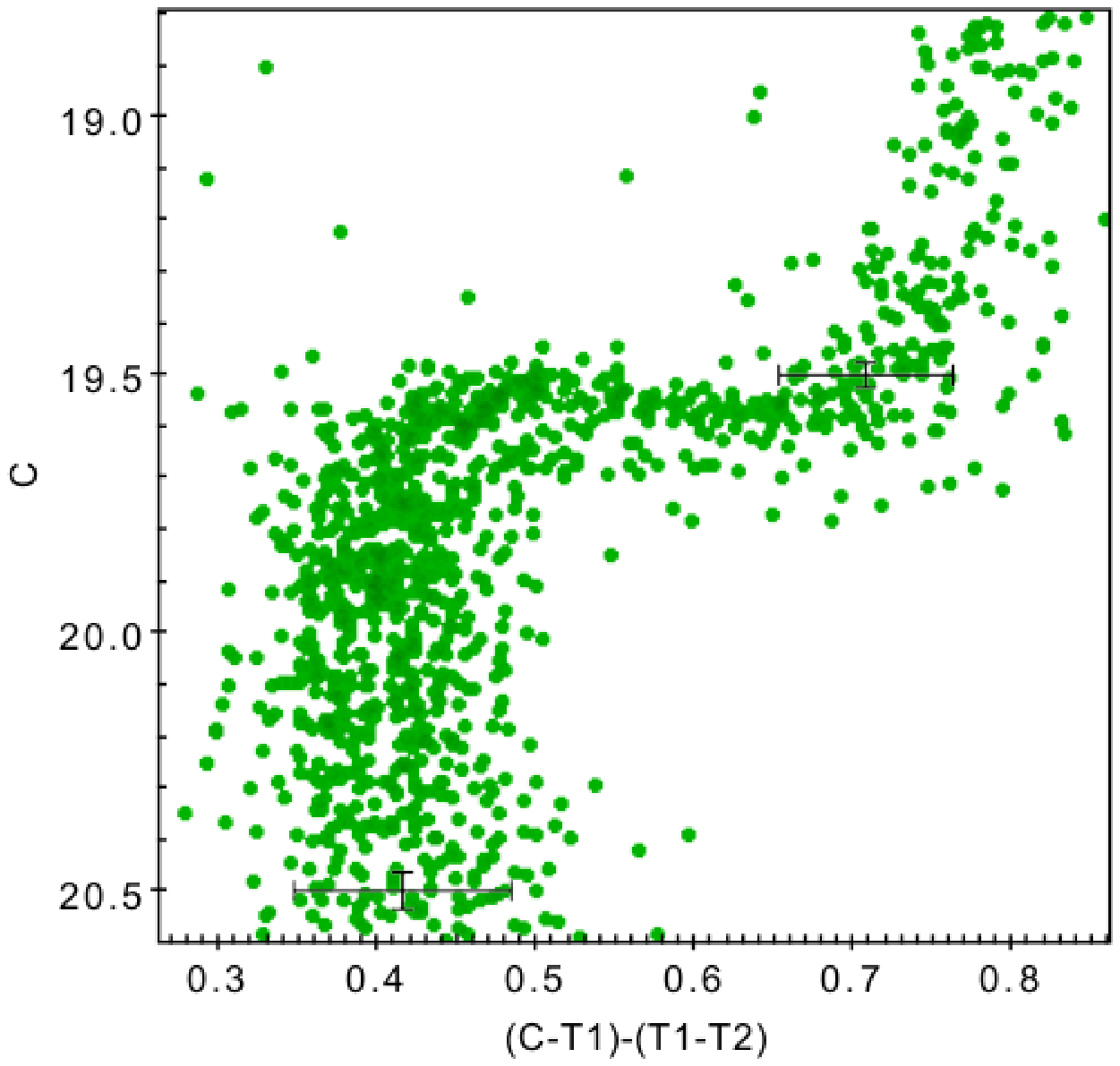}
  	\includegraphics[width=0.495\textwidth]{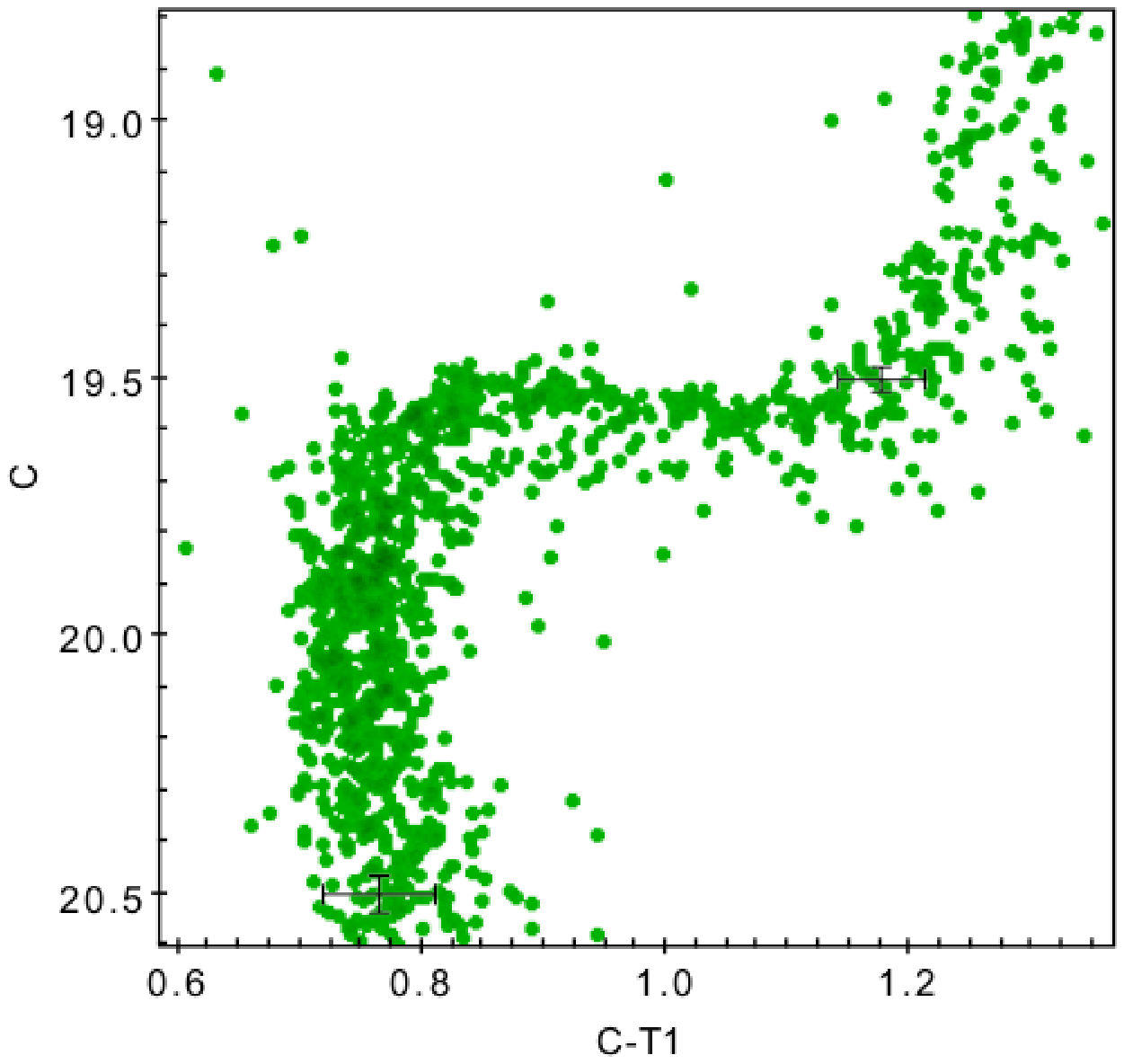}
  	\caption{Comparison of the lower RGB and SGBs of NGC1851 seen in $C,T_1,T_2$ vs C (top) and $C-T_1$ vs C (bottom).}
    \label{SGB}
\end{figure}

Figure \ref{SGB} shows a comparison of the lower RGB and SGBs of NGC1851 between $C,T_1,T_2$ vs C (top) and $C-T_1$ vs C (bottom).
Subsets were taken from both colors trying to cover all the SGB. The fainter SGB is somewhat more visible in the former.

Radial distributions between the Bright-SGB and Faint-SGB were compared in both subsets. The K-S test in $C-T_1$ give a p-value of 0.729 while in $C,T_1,T_2$ give 0.590. Both subsets show no significant differences in radial distributions between Bright-SGB and Faint-SGB stars. This is in agreement with \citet{Milone2009} who also did not find differences in the radial distributions of the SGBs of NGC 1851. 

The combined samples of the SGB and RGB from each Bright-SGB/Blue-RGB and Faint-SGB/Red-RGB were analyzed to discard the possibility that the lack of differences in the radial distributions is due to the low amount of stars. $C-T_1$ give a p-value of 0.095 while $C,T_1,T_2$ give 0.128. Both of these values are considered values too high to assure that there are differences in radial distributions. Recall that indeed C14 found differences in the radial distributions of the MS but none in the RGB and HB, even after combining them. Also, the percentage of Red-RGB/Faint-SGB stars is 14.5\%$\pm$2.5\% in $C-T_1$ and 14,1\%$\pm$3.3\% in $C,T_1,T_2$, confirming that there is no significant improvement with $C,T_1,T_2$ with respect to $C-T_1$ for this case. For this case we can also see that the percentage of error in $C,T_1,T_2$ is higher than in $C-T_1$ due to, again, the fact that $C,T_1,T_2$ does not help to split the secuences\\

\emph{-First and second Populations in the Blue-RGB}\\
According to \citet{Campbell2012}, both Blue-RGB and Red-RGB could possess a first and second population of stars. This idea is also supported by \citet{Milone2017}, who found two 2P sequences in the chromosome map of NGC 1851, with a hint of a second 1P as well.\\
For our case, the number of stars in the Red-RGB is too small to analyze it in search of a 1P/2P division, so we will only analyze the Blue-RGB.

\begin{figure*}
\centering
  	\includegraphics[width=0.32\linewidth]{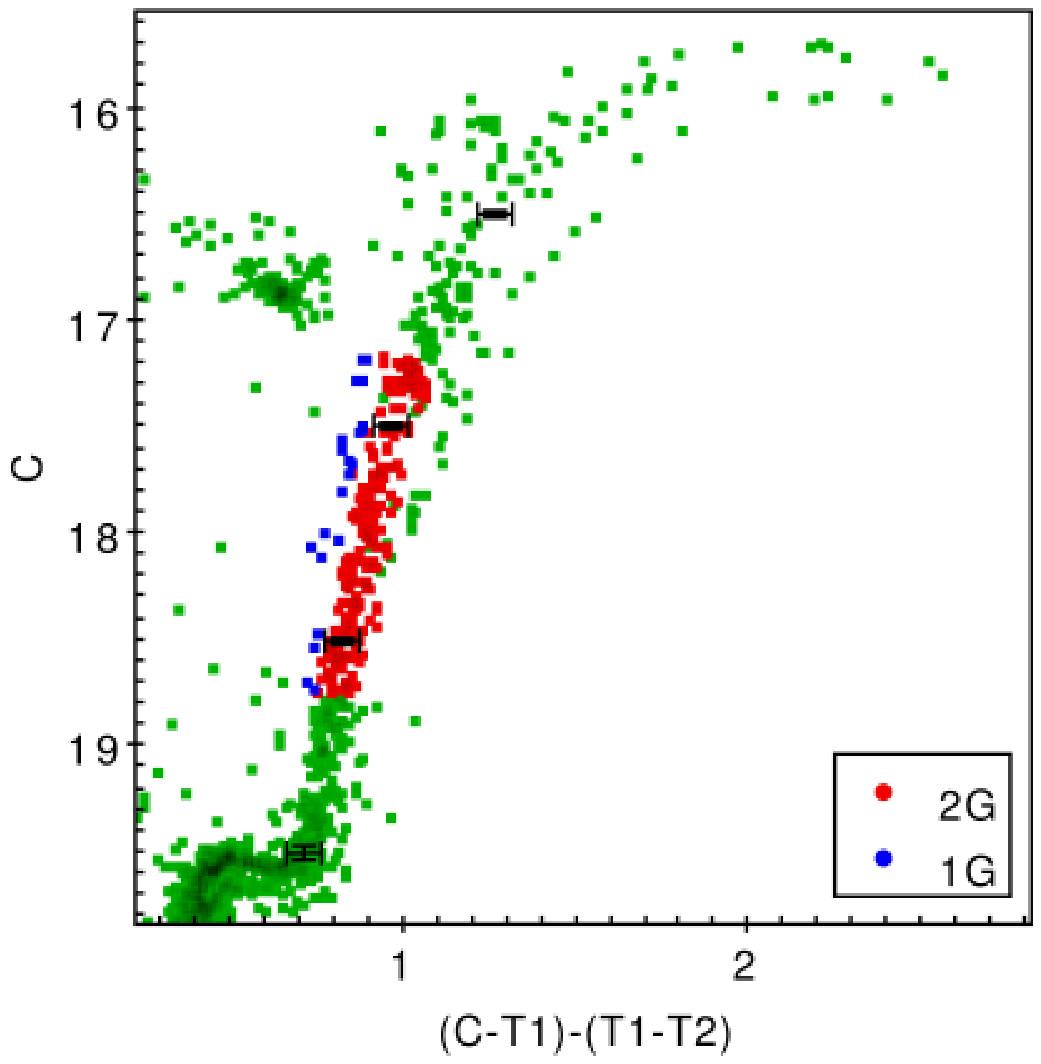}
  	\includegraphics[width=0.32\linewidth]{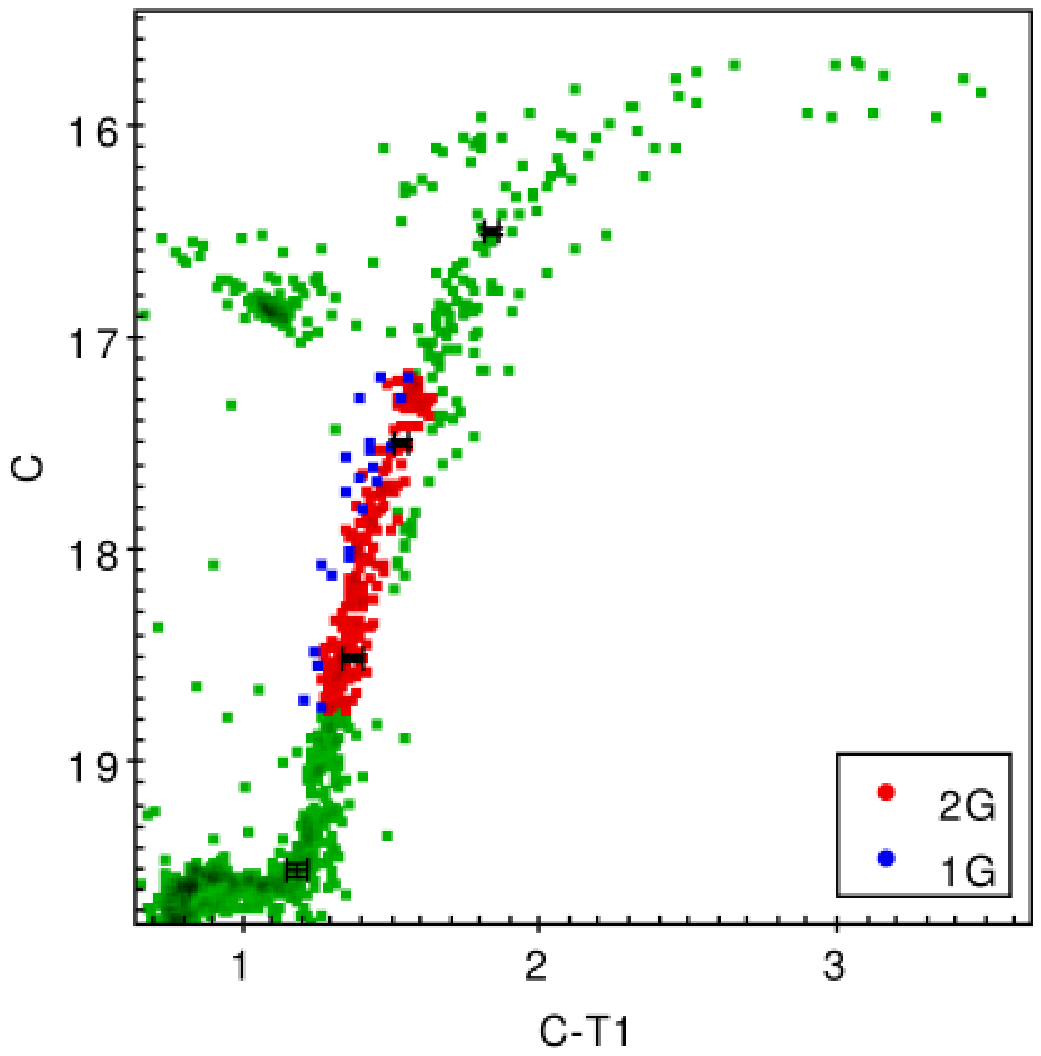}
  	\includegraphics[width=0.32\linewidth]{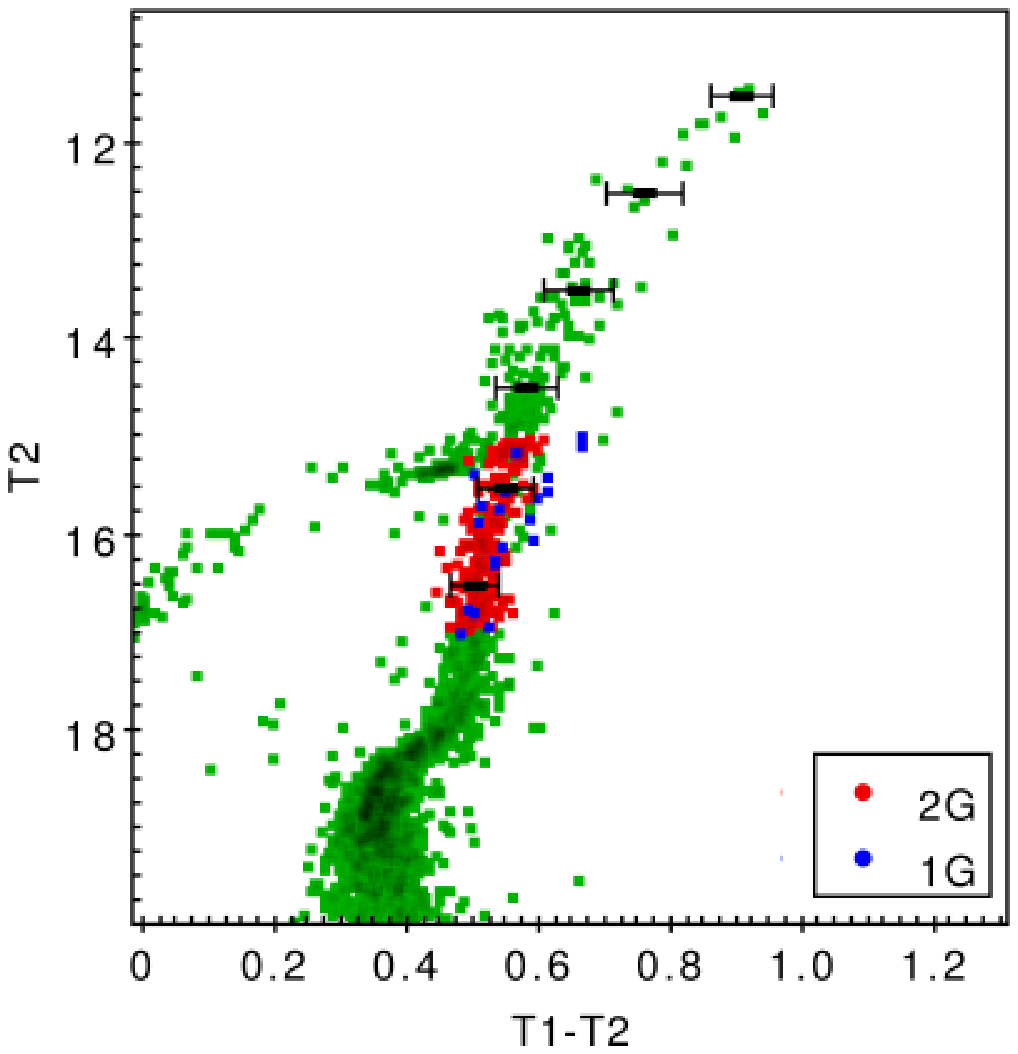}
  	
  	\includegraphics[width=0.32\linewidth]{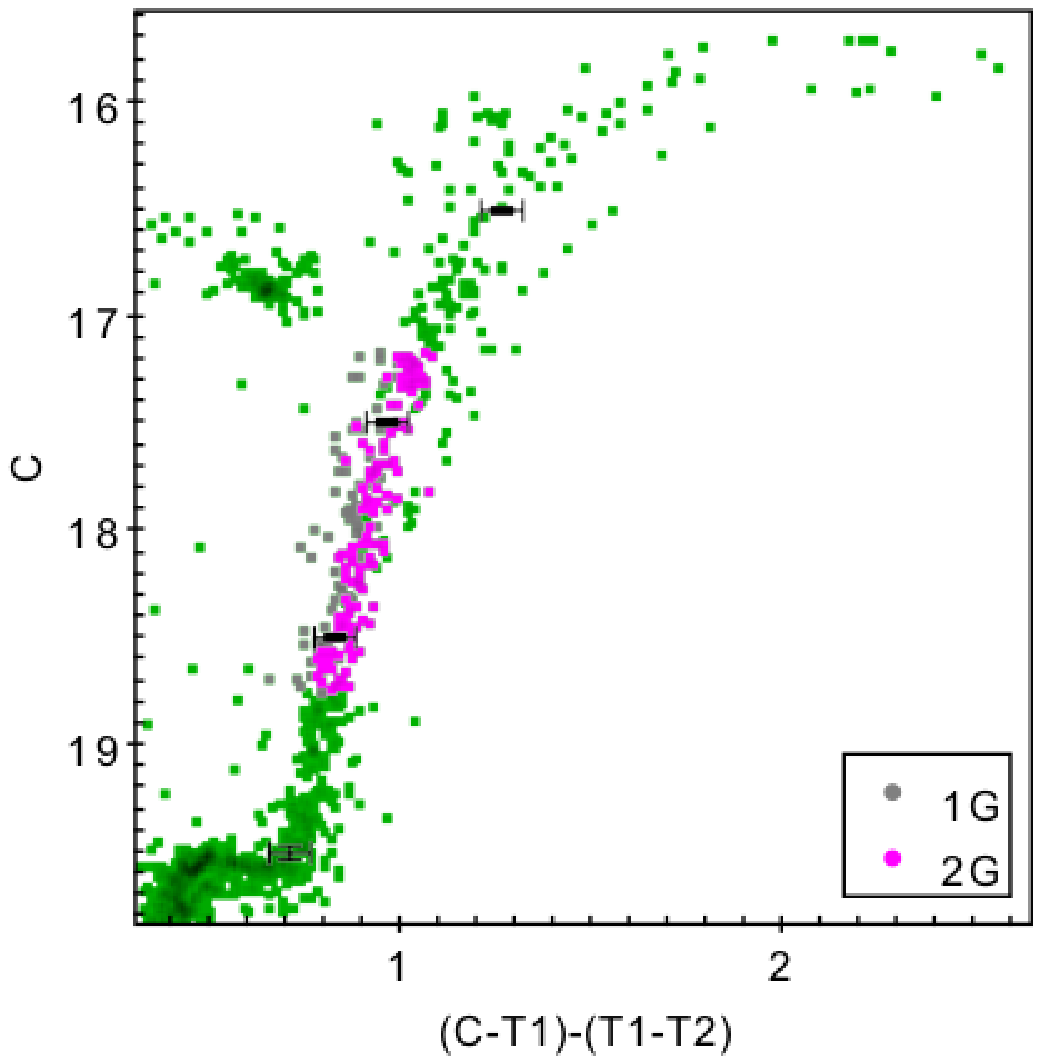}
  	\includegraphics[width=0.32\linewidth]{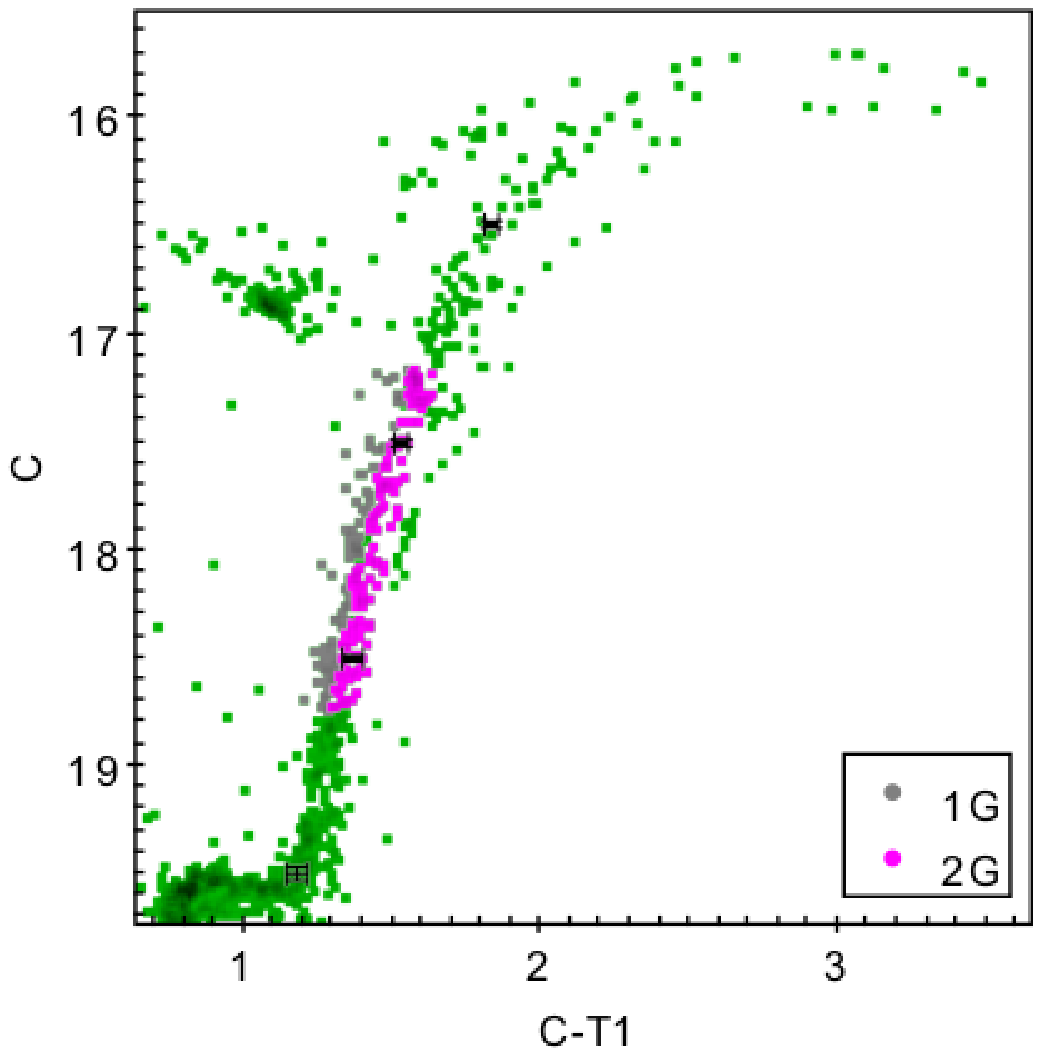}
  	\includegraphics[width=0.32\linewidth]{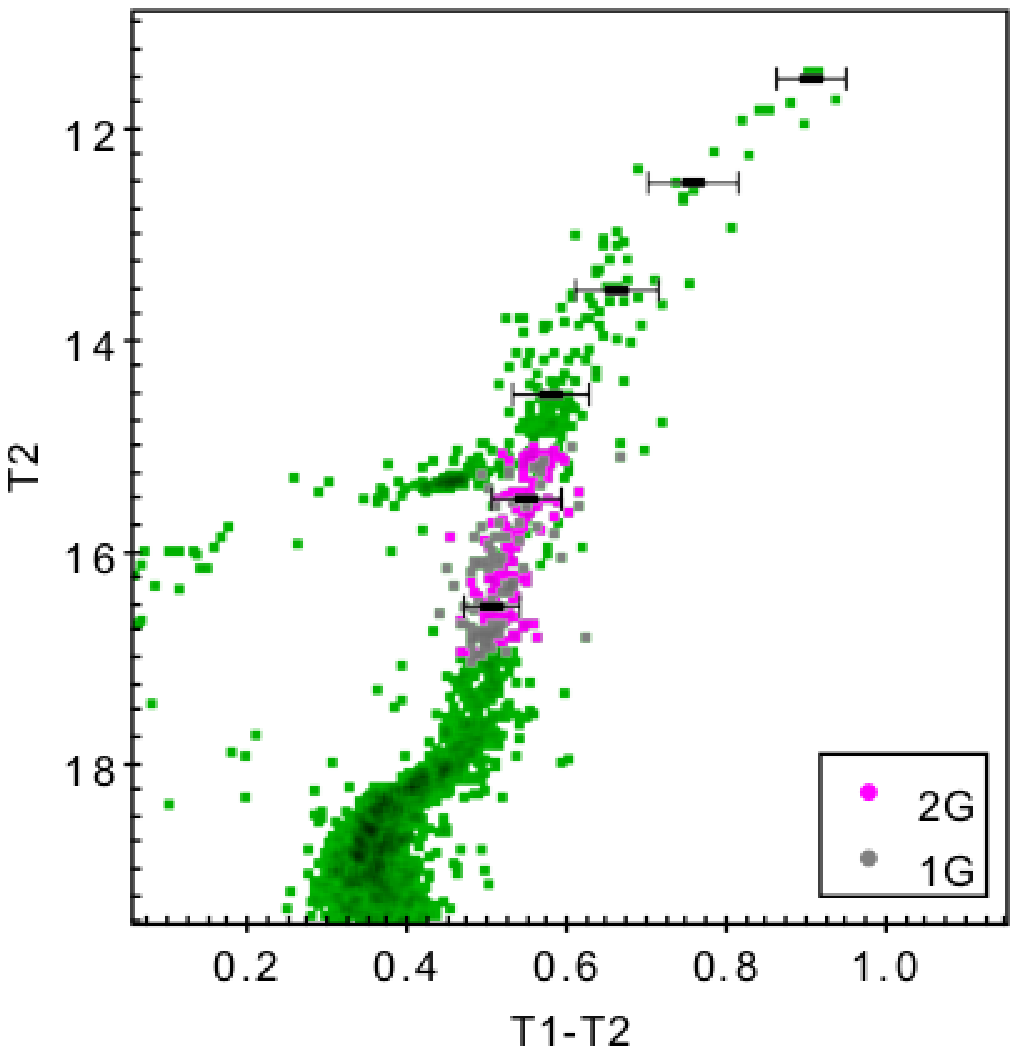}
  	
  	\includegraphics[width=0.32\linewidth]{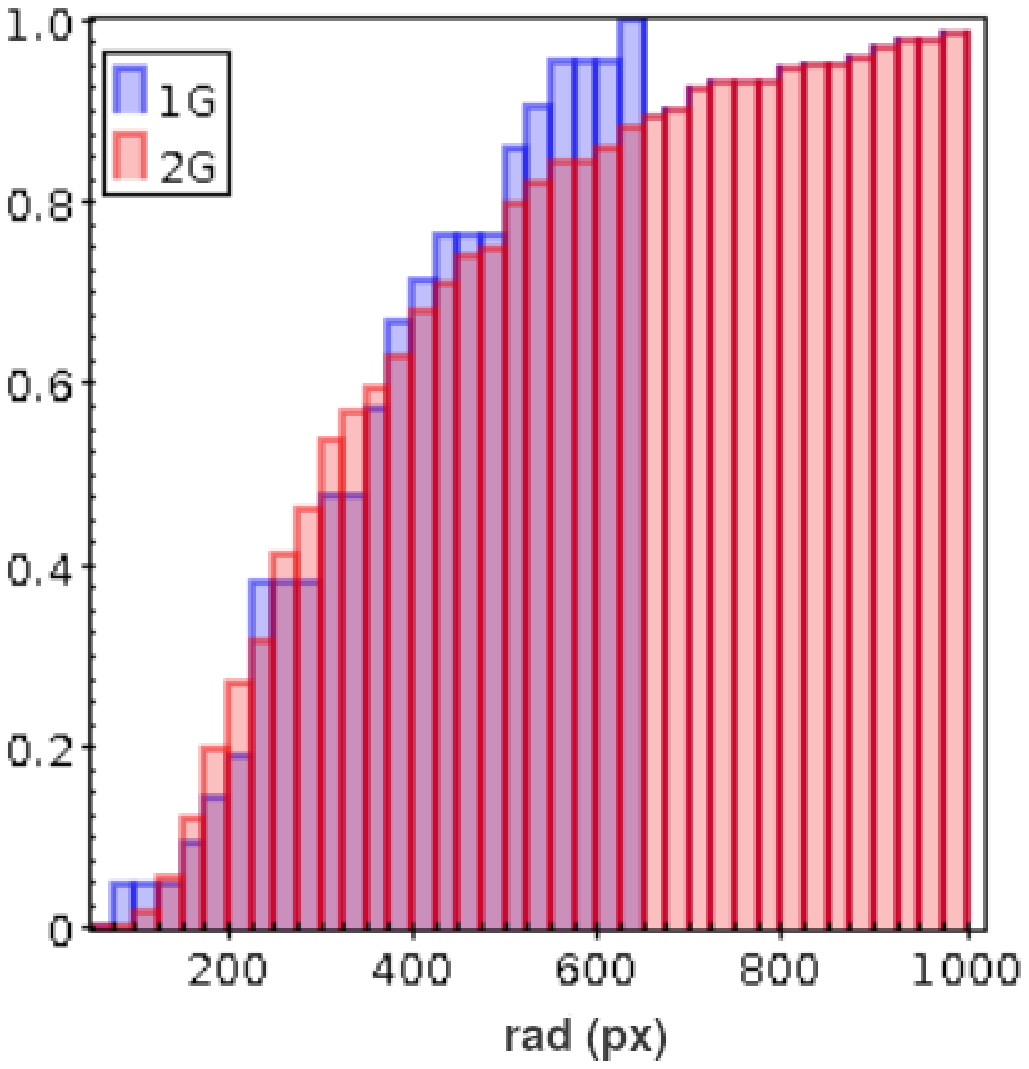}
  	\includegraphics[width=0.32\linewidth]{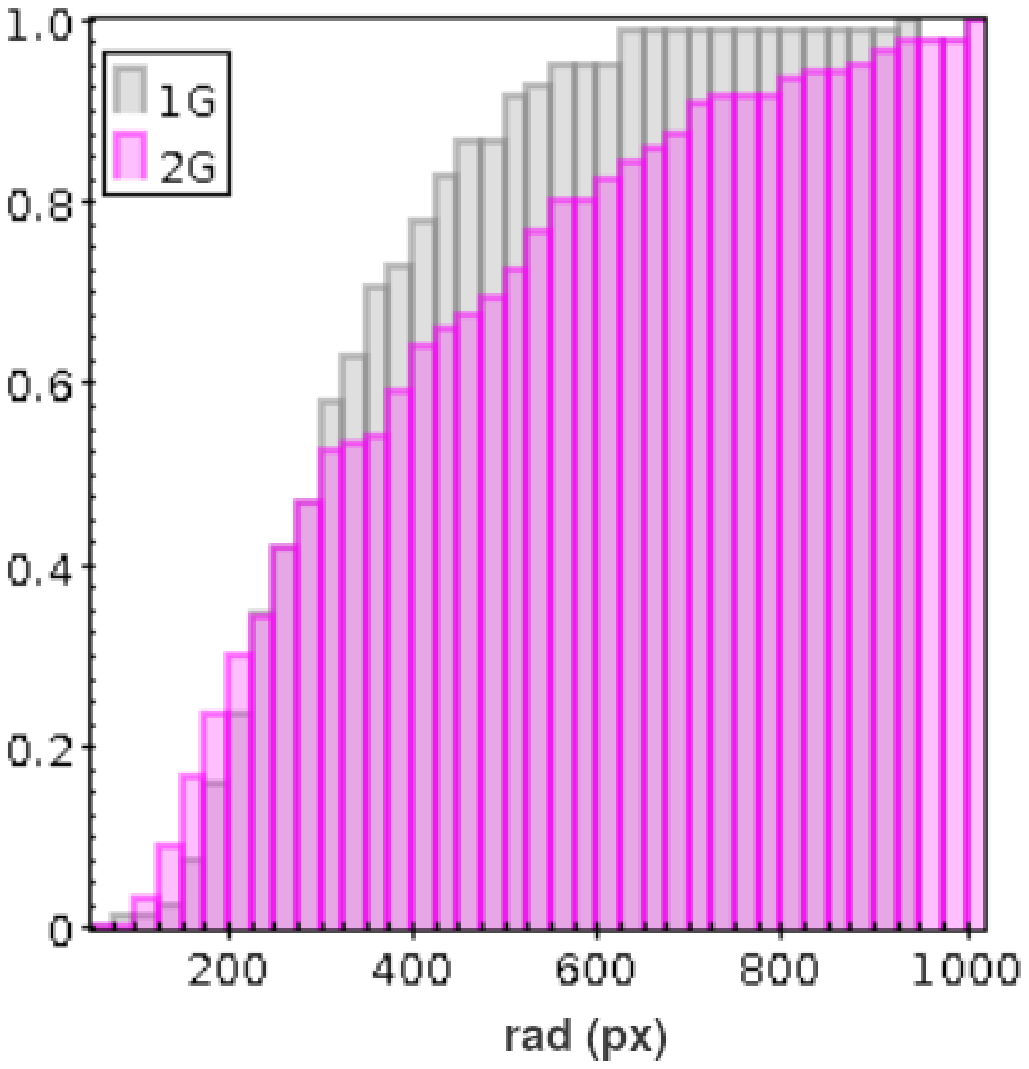}
   	\includegraphics[width=0.32\linewidth]{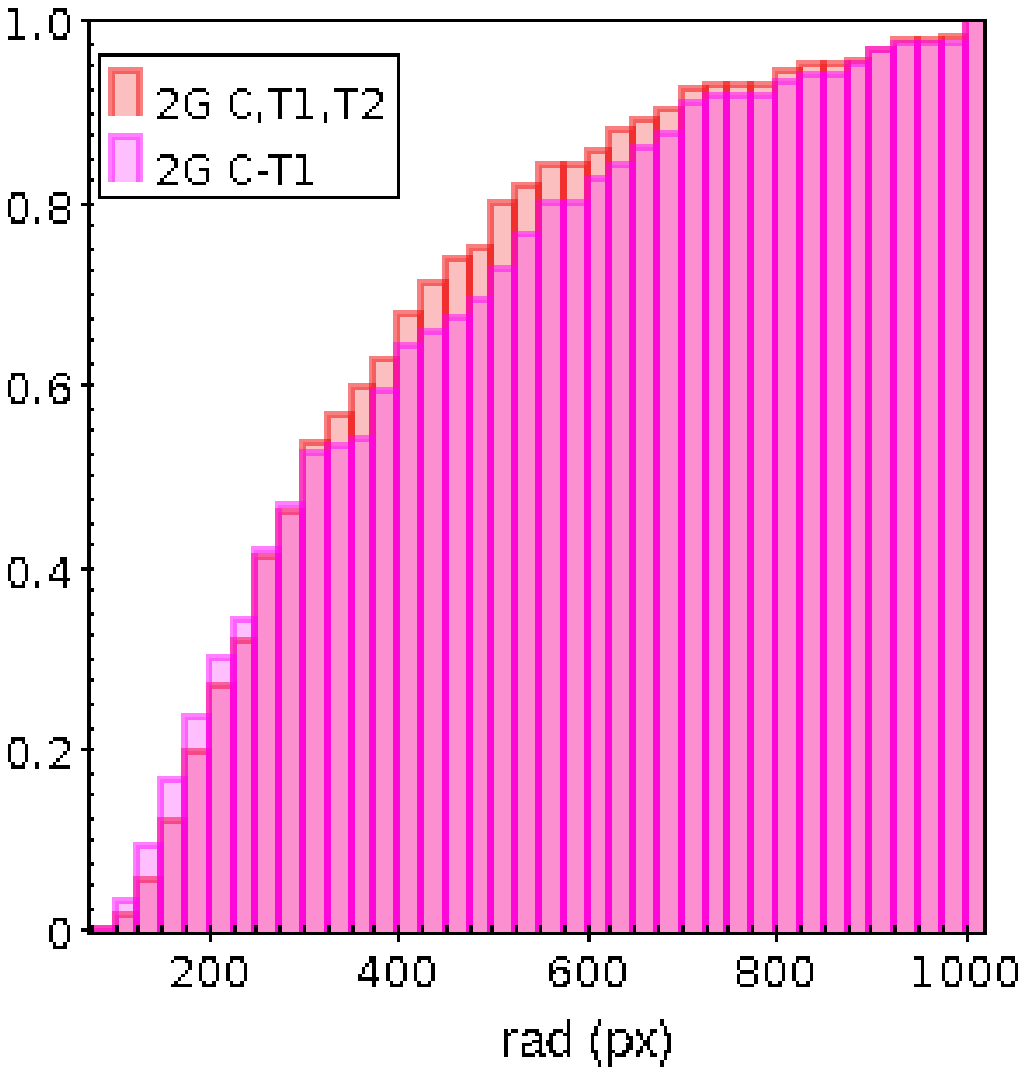}

    \caption{NGC 1851 Blue-RGB: Upper panels: A 1P/2P subset chosen from the color C,T1,T2 shown in C-T1 and T1-T2.
    Mid panels: A 1P/2P subset chosen from the color C-T1 shown in C,T1,T2 and T1-T2.
    Bottom panels: Left: Radial distributions of 1P and 2P of the subset from C,T1,T2. Middle: Radial distributions of 1P and 2P of the subset from C-T1. Right: Comparison of the 2P from the C,T1,T2 and the 2P from C-T1.}
    \label{18511P2P }
\end{figure*}

Following the same procedure used in NGC 7099, we used the errorbars from each color to separate the 1P from the 2P. We found a small fraction of stars lying at the left side of the Blue-RGB in $C-T_1$ and $C,T_1,T_2$, deemed as 1P stars. Figure \ref{18511P2P } shows the comparison of the 1P subset chosen from $C,T_1,T_2$ (upper panels) and $C-T_1$ (central panels). This time the chosen subset from $C,T_1,T_2$ follow the expected behaviour for a different population: well defined at one side of the RGB in $C,T_1,T_2$, partially less defined at the same side of the RGB in $C-T_1$, and even less defined and at the opposite side of the RGB in $T_1-T_2$, although the last point is not as clear in $C-T_1$. With $C,T_1,T_2$ the Blue-RGB has 10.3\%$\pm$19.5\% of 1P stars while with $C-T_1$ the amount of 1P stars is 40.3\%$\pm$19.8\%. 

The radial distributions in $C-T_1$ gives a P-value of 0.014 while $C,T_1,T_2$ show a P-value of 0.870, although the last result is not as reliable since the 1P are only 21 stars.

 \begin{table}
 \centering
 \caption{Comparison error vs width in the Blue-RGB of NGC 1851.} \label{1851vs2}
 \begin{tabular}{llll}\toprule
 Magnitude Range & Mean width & Mean error & Ratio\\\midrule
 & C-$T_1$ vs C\\\midrule
 17.2-18 & 0.042 & 0.023 & 1.83 \\   
 18-18.8 & 0.042 & 0.024 & 1.75 \\\midrule
 & C,$T_1$,$T_2$ vs C \\\midrule
 17.2-18 & 0.044 & 0.047 & 0.93\\
 18-18.8 & 0.039 & 0.039 & 1.00\\\bottomrule
  \end{tabular}
  \end{table} 

A width to error ratio analysis indicates that $C-T_1$ has a mean ratio of 1.79 while C,$T_1$,$T_2$ vs C has a mean ratio of 0.97. $C-T_1$ ratio is too small to confirm or reject the presence of MPs while the width of C,$T_1$,$T_2$ is completely due to errors ($T_1-T_2$ was not considered in this table since the Red-RGB is inside the Blue-RGB, hence the values of the last are the same of those in table \ref{1851vs}).

\begin{figure}
 	\includegraphics[width=0.48\textwidth]{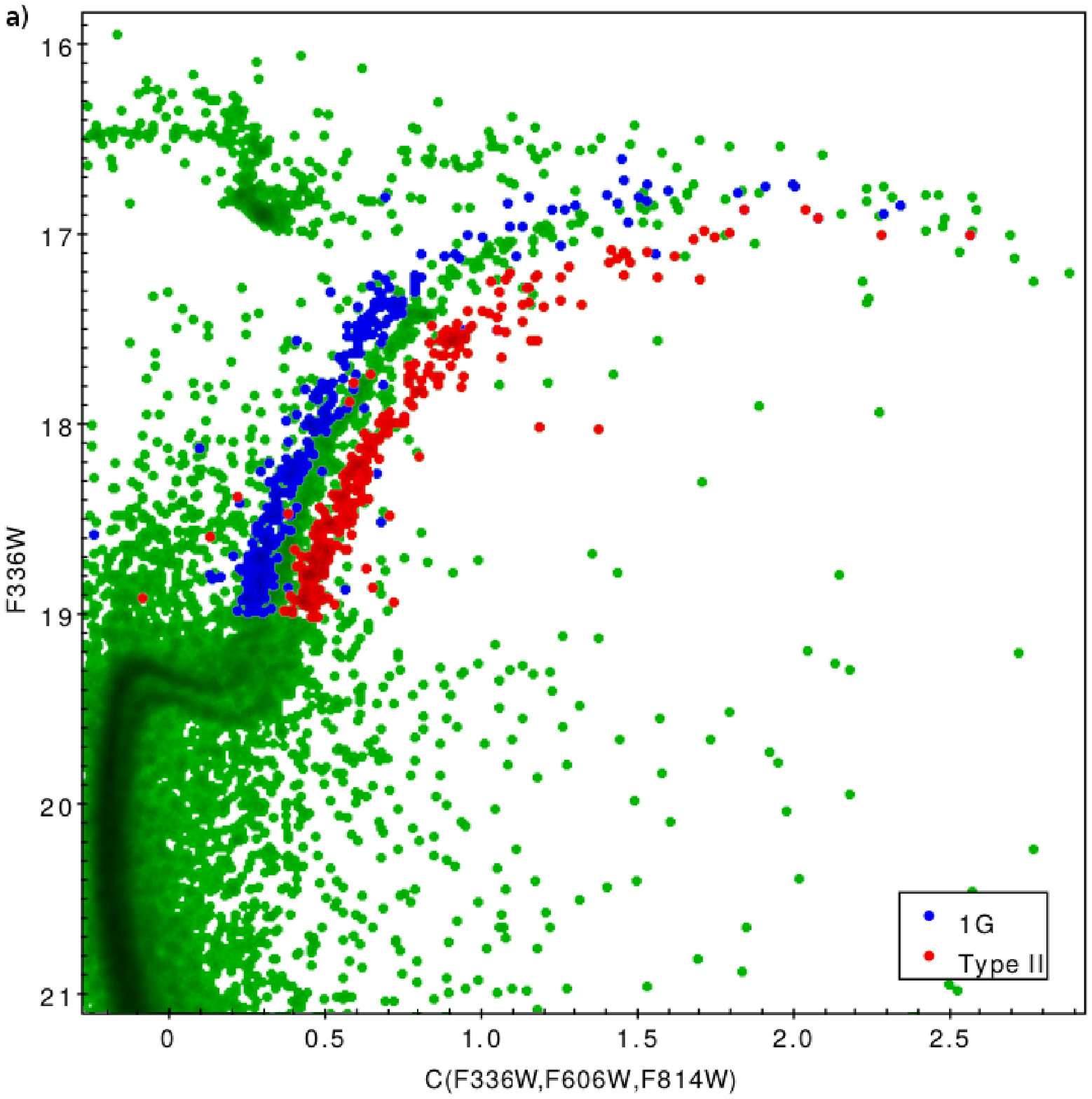}
  	\includegraphics[width=0.51\textwidth]{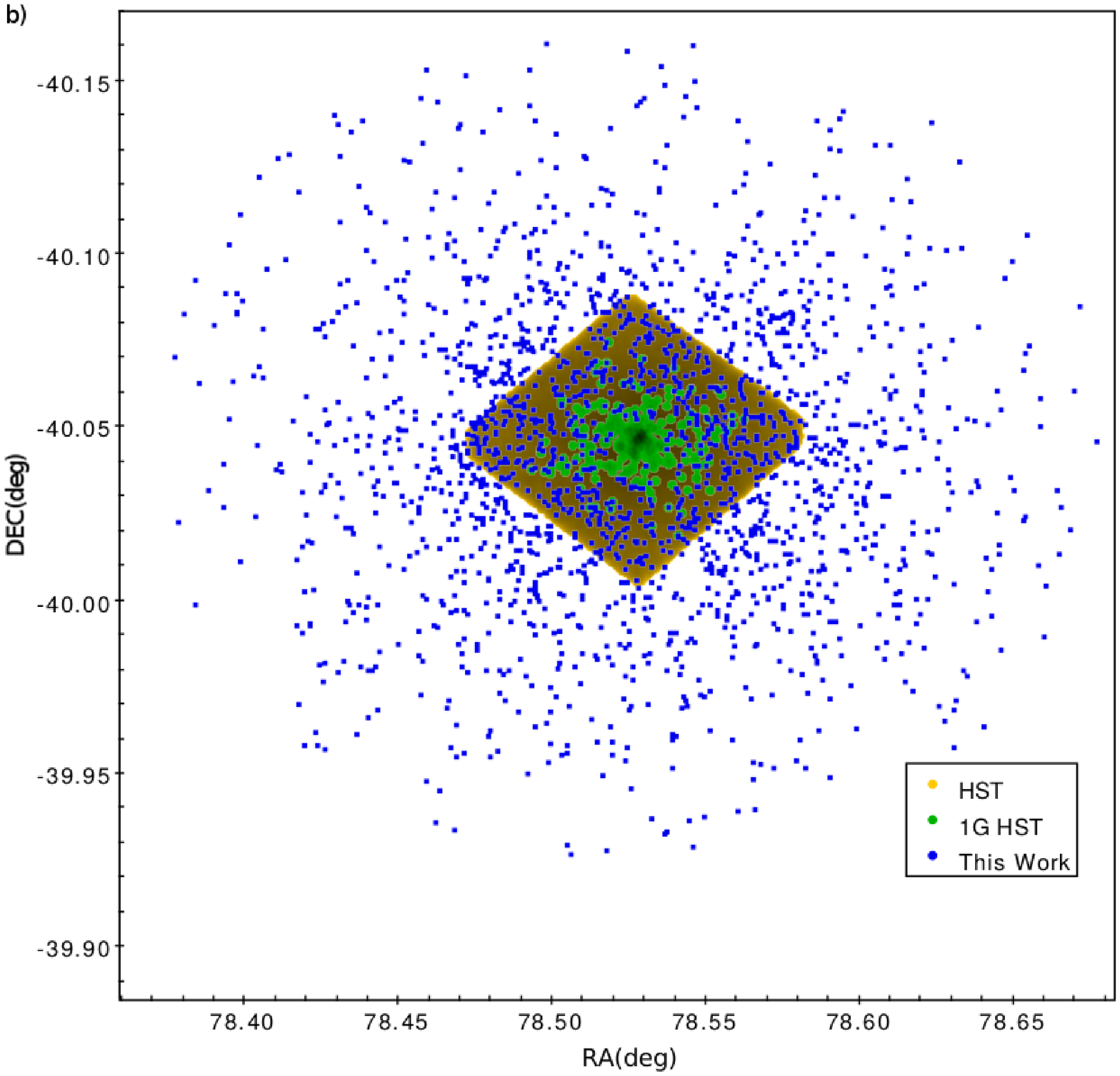}
  	\caption[NGC 1851: 1P using HST and radial distributions]{NGC 1851: a)CMD using F336W, F606W and F814W from HST in replacement of C, $T_1$ and $T_2$. The blue and red dots are the 1P and Red-RGB stars respectively, identified by Milone et al. (2017). b)Spatial distributions of the stars of our work and HST. Most of the 1P stars (green dots) were removed in our work due to the crowding.}
    \label{1851magic}
\end{figure}

A further analysis was realized using data from the HST UV Globular cluster Survey described in \citet{Piotto2015}, in an attempt to verify if our subset chosen as 1P in the Blue-RGB of NGC 1851 was correct or not. Taking as 1P the subset chosen in \citet{Milone2017} using the "Magic Trio" in NGC1851 we recreated our Washington Trio using the Filters F336W, F606W and F814W in replacement of C, $T_1$ and $T_2$ respectively. We obtained a very similar CMD, as shown in Figure \ref{1851magic}a, where the blue stars are the 1P, the green RGB are the 2P stars and the red stars are the Red-RGB stars. The reason for the small amount of 1P stars is because most of them lie less than 22 arcseconds from the center, and we cut those stars (up to 21.75 arcseconds of radius) due to the crowding, as seen in Figure \ref{1851magic}b. This would explain the small width to error ratios and the high percentage of error respect to the amount of 1P stars.
 Despite this, we were able to detect a small amount of 1P stars thanks to the $C,T_1,T_2$ color, confirming it as a complementary method to detect MPs.

\section{Conclusions}

In this work we have improved the Washington photometry of the clusters NGC 7099 and NGC 1851 done in \citet[F17]{Frelijj2017} and \citet[C14]{Cummings2014}, respectively. We have determined the optimum way to reduce the data, thus obtaining the highest number of stars with the minimum possible errors. We also added PM information from Gaia to help select members. Finally, we created a new color combination $(C-T_1)-(T_1-T_2)$ and we tested it in two clusters.
\newline
We conclude that:

1) An expected behaviour for a 1P/2P subset is to be at one side of the RGB in $C,T_1,T_2$; at the same side but less defined in $C-T_1$; and even less defined, at the opposite side, in $T_1-T_2$.

2)The subsets chosen as 1P and 2P in F17 for NGC 7099 are incorrect, since the stars belonging to the 1P are actually field stars, removed now thanks to the PM provided by the Gaia mission. This explains why we got a p-value of 0.0 and the 1P stars having radial distributions more concentrated to the center than the 2P.

3)The new color combination widens the RGB of NGC 7099 better than $(C-T_1)$, allowing to properly select the Population subsets. However, $(C-T_1)$ still has the best width/error ratio. Depending on the criteria used, $C,T_1,T_2$ would have a stronger central concentration than $(C-T_1)$ or weaker. While the 2P subset chosen from $C-T_1$ has the highest fraction of stars within $\sim$300px from the center, the subset extends until $\sim$980px, while the 2P subset from $C,T_1,T_2$ has no stars beyond $\sim$780px from the center.

4)We find a very small number of 1P stars at the left side of the Blue-RGB of NGC 1851 using both $C,T_1,T_2$ and $C-T_1$. Despite the subset of $C,T_1,T_2$ being slightly more accurate, the small number of stars complicates any study. A comparison with analogous HST filters (F336W,F606W and F814W) shows a very similar CMD with a higher amount of 1P stars at the same side of our small subset, confirming our findings. But a spatial analysis of those stars shows that most of them were removed in our catalog due to the crowding of stars at the center. 

5) The Red-RGB in NGC 1851 does not follow the expected behaviour for a common 2P group of stars. Instead, it appears at the same side, without any improvement, in both $C,T_1,T_2$ and $C-T_1$ colors, while in $T_1-T_2$ they are completely merged with the stars from the Blue-RGB. Also, little or no difference is seen in the radial distributions between the stars of the Blue-RGB and Red-RGB using the old and new color combinations, even when combining the samples with the bright-SGB and faint-SGB, respectively. 

6)- The 2P percentage in the RGB of NGC 7099 is 23.2\%$\pm$25\% for $C-T_1$ and 44.9\%$\pm$16.7\% for $C,T_1,T_2$.\\
  - The percentage of Faint-SGB/Red-RGB stars respect to the total number of stars in the SGBs and RGBs in NGC 1851 is 14.5\%$\pm$2.5\% for $C-T_1$ and 14.1\%$\pm$3.3\% for $C,T_1,T_2$.\\
  - The percentage of 1P stars in the Blue-RGB of NGC 1851 is 40.3\%$\pm$19.8\% for $C-T_1$ and 10.3\%$\pm$19.5\% for $C,T_1,T_2$.\\
  Comparing these percentages with those of Milone et al.(2017)($\sim$62\% of 2P stars in NGC 7099, $\sim$3\% of Red-RGB stars in NGC 1851 and $\sim$26.4\% of 1P stars in NGC 1851) we find little relation. This difference might occur not only due to our high percentages of error in the populations, but also because we removed the center of our cluster (a radius of 21,75'' in NGC 1851 and 34.8'' in NGC 7099) in order to avoid issues due to the crowding. In addition, the HST photometry used here covers the field up to a radius of only $\sim$2.47' while our photometry covers up to a radius of $\sim$7.25'. 

  This means our $C,T_1,T_2$ color combination is a reliable method to detect MPs since it improve the detection of MPs and decreases the uncertainties of the defined 1P and 2P. But looking at the width that produces compared to the increase of errors that this implies is not as effective as $C-T_1$. It is also less efficient, as the latter requires only observations in 2 filters. So we recommend its use to detect MPs, but as a complementary method together with $C-T_1$ and $T_1-T_2$ separately.

\section*{Acknowledgements}
We thank the anonymous referee for his/her suggestions and comments that have improved the quality of the paper.\\
H.F. acknowledges financial support from Agencia Nacional de Investigacion y Desarrollo (ANID) grant 21181653.\\
D.G. and S.V. gratefully acknowledge support from the Chilean Centro de Excelencia en Astrof\'isica
y Tecnolog\'ias Afines (CATA) BASAL grant AFB-170002.\\
D.G. also acknowledges financial support from the Direcci\'on de Investigaci\'on y Desarrollo de
la Universidad de La Serena through the Programa de Incentivo a la Investigaci\'on de
Acad\'emicos (PIA-DIDULS).\\
C.M. thanks the support provided by FONDECYT Postdoctorado No.3210144.\\
S.V. gratefully acknowledges support by the ANID BASAL project ACE210002.\\
S.V. gratefully acknowledges support by the ANID BASAL projects ACE210002 and FB210003.\\
S.V. gratefully acknowledges the support provided by Fondecyt reg. n. 1220264.\\

\section*{Data Availability}
This work has made use of data from the European Space Agency (ESA) mission
{\it Gaia} (\url{https://www.cosmos.esa.int/gaia}), processed by the {\it Gaia}
Data Processing and Analysis Consortium (DPAC, \url{https://www.cosmos.esa.int/web/gaia/dpac/consortium}). Funding for the DPAC has been provided by national institutions, in particular the institutions
participating in the {\it Gaia} Multilateral Agreement.

This work has made use of data from the HST UV Globular cluster Survey \url{http://groups.dfa.unipd.it/ESPG/treasury.php}.

The Photometric raw data from SOAR and Swope telescopes analysed in this article will be shared on reasonable request to the corresponding author (HF).

\end{document}